\documentclass[aps,pra,twocolumn,groupedaddress]{revtex4-1}
\usepackage{hyperref}

\usepackage{graphicx}
\usepackage{latexsym}

\usepackage{bm,array}

\usepackage{amsfonts}
\usepackage{amssymb}
\usepackage{amsmath}
\usepackage{upgreek}

\usepackage{tabstackengine}
\stackMath

\newcommand{\mtx}[2]{\left(\begin{array}{#1}#2\end{array}\right)}

\newcolumntype{x}[1]{
>{\centering\hspace{0pt}}p{#1}}

\begin{document}

% Use the \preprint command to place your local institutional report
% number in the upper righthand corner of the title page in preprint mode.
% Multiple \preprint commands are allowed.
% Use the 'preprintnumbers' class option to override journal defaults
% to display numbers if necessary
%\preprint{}

%Title of paper
\title{Interpreting symplectic linear transformations in a two-qubit phase space}

% repeat the \author .. \affiliation  etc. as needed
% \email, \thanks, \homepage, \altaffiliation all apply to the current
% author. Explanatory text should go in the []'s, actual e-mail
% address or url should go in the {}'s for \email and \homepage.
% Please use the appropriate macro foreach each type of information

% \affiliation command applies to all authors since the last
% \affiliation command. The \affiliation command should follow the
% other information
% \affiliation can be followed by \email, \homepage, \thanks as well.
\author{William K.~Wootters}
%\email[]{Your e-mail address}
%\homepage[]{Your web page}
%\thanks{}
%\altaffiliation{}
\affiliation{Department of Physics, Williams College, Williamstown, Massachusetts 01267, USA}

%Collaboration name if desired (requires use of superscriptaddress
%option in \documentclass). \noaffiliation is required (may also be
%used with the \author command).
%\collaboration can be followed by \email, \homepage, \thanks as well.
%\collaboration{}
%\noaffiliation

%\date{\today}

\begin{abstract}
% insert abstract here
For the continuous Wigner function and for certain discrete Wigner functions,
permuting the values of the Wigner function in accordance with a symplectic linear transformation is equivalent to performing a certain
unitary transformation on the state.  That is, performing this unitary transformation is simply a matter of moving
Wigner-function values around in phase space.   This result holds in particular for the simplest discrete Wigner function defined on a $d \times d$ phase space when the Hilbert-space dimension $d$ is odd.  
It does not hold for a $d \times d$ phase space if the dimension is even.  Here we show, though, that a 
generalized version of this correspondence does apply in the case of a two-qubit phase space.  In this case, a symplectic linear permutation of the points of the phase space,
together with a certain reinterpretation of the Wigner function, is equivalent to a unitary
transformation.  
\end{abstract}

% insert suggested PACS numberis s in braces on next line
\pacs{}
% insert suggested keywords - APS authors don't need to do this
%\keywords{}

%\maketitle must follow title, authors, abstract, \pacs, and \keywords
\maketitle

% body of paper here - Use proper section commands
% References should be done using the \cite, \ref, and \label commands
%\section{}
% Put \label in argument of \section for cross-referencing
%\section{\label{}}
%\subsection{}
%\subsubsection{}

\section{Introduction}	%) A SECTION HEADING

In classical mechanics, the concept of a phase space, with dimensions representing coordinates and momenta, is utterly central.  While it is less central in quantum mechanics,
phase-space representations of quantum states such as the Wigner function have proved fruitful in a wide variety of applications \cite{Hillery,Weinbub}.  
A particular point of contact between classical phase space and quantum phase space is the role played by symplectic transformations.
In the quantum case, a symplectic linear transformation of the phase space corresponds to a unitary transformation in the Hilbert space \cite{Arvind,Ferraro}.  For example,
to perform a unitary squeezing operation on the Wigner function representing a quantum state, one simply squeezes the Wigner function itself: its
values are moved around in phase space in accordance with the appropriate linear transformation.  In the case of classical mechanics,
a canonical transformation is typically not linear but its Jacobian at every point is a symplectic matrix.  

To a certain extent, the strong connection between symplectic linear transformations in phase space and a particular class of unitary transformations in Hilbert space carries over to the 
discrete Wigner function, which is used to represent states of a system with a finite-dimensional Hilbert space.  The discrete phase space we consider in this paper is a two-dimensional vector space over a finite field \cite{Gibbons,Klimov,Vourdas,Gross}.  
Such a phase space, when the field has $d$ elements, is appropriate for representing a system with a $d$-dimensional Hilbert space.  
If $d$ is odd, it is possible to define on this phase space, essentially uniquely, a discrete Wigner function that exhibits the same kind of correspondence between symplectic
transformations and unitary transformations that one sees in the continuous Wigner function \cite{Klimov,Vourdas,Gross,Bengtsson,Zhu3}.  That is, if $L$ is a symplectic linear transformation of the phase 
space---in our setting this simply means that $L$ can be represented by a unit-determinant $2 \times 2$ matrix with entries in the finite field---then there exists
a unitary transformation $U_L$ corresponding to $L$ such that 
\begin{equation}  \label{Wignermove}
W_\alpha(U_L \rho U_L^\dag) = W_{L^{-1}\alpha}(\rho).
\end{equation}
Here $W_\alpha(\rho)$ is the Wigner function representing the density matrix $\rho$, evaluated at the point $\alpha$ in the discrete phase space.  
Thus, in order to find the value of the post-transformation Wigner function at the point $\alpha$, one simply
asks what point in phase space was sent to $\alpha$ by $L$.  The value of the pre-transformation Wigner function at that point, $L^{-1}\alpha$, has now been
moved to the point $\alpha$.

The connection (\ref{Wignermove}) between symplectic linear transformations and unitary transformations does not hold for a $d \times d$ phase space when the dimension $d$ is a power of 2.  This is unfortunate, since 
systems of qubits are of great interest in quantum information science.  

We can see how Eq.~(\ref{Wignermove}) fails in the case of a single qubit.  
In that case, we adopt the following definition of the discrete Wigner function \cite{Feynman,Wootters}.  
First, the phase-space points are labeled by $\alpha = (\alpha_q, \alpha_p)$, where $\alpha_q$ and $\alpha_p$ take values in the two-element field ${\mathbb F}_2 = \{0,1\}$.  (We actually treat $\alpha$ as a column vector---a matrix acting on it acts from the left---but we sometimes write it as a row vector to make the
typesetting less awkward.)  We picture $\alpha_q$ as the horizontal coordinate and $\alpha_p$ as the vertical coordinate, and we picture the origin, 
$\alpha = (0,0)$, in the lower left-hand corner of the $2 \times 2$ grid.  For the qubit density matrix $\rho$, the corresponding Wigner function is defined as
\begin{equation}  \label{Wdef}
W_\alpha(\rho) = \tfrac{1}{2} \hbox{Tr} (A_\alpha \rho),
\end{equation}
where the phase point operators $A_\alpha$ are given by
\begin{equation} \label{A00}
A_{(0,0)} =  \tfrac{1}{2}[I +X + Y +Z]
\end{equation}
and 
\begin{equation} \label{trans}
\begin{split}
&A_{(1,0)} = X A_{(0,0)} X, \\ 
&A_{(1,1)} = Y A_{(0,0)} Y, \\
&A_{(0,1)} = Z A_{(0,0)} Z.
\end{split}
\end{equation}
Here $X$, $Y$, and $Z$ are the Pauli matrices.  Note that the $A$'s constitute an orthogonal basis for the real vector space of Hermitian $2 \times 2$ matrices:
$\hbox{Tr}(A_\alpha A_\beta) = 2 \delta_{\alpha\beta}$.
With this definition, the property of the Wigner function expressed in Eq.~(\ref{Wignermove}) is equivalent to the following property of the phase point operators:
\begin{equation} \label{Amove}
U_L A_\alpha U^\dag_L = A_{L\alpha}.
\end{equation}

There are exactly six symplectic matrices $L$ for this simple phase space.  Three of the six are of the form $R^s$, with $s=0, 1, 2$, where
$R$ is the ``rotation'' matrix
\begin{equation}
R = \mtx{cc}{1 & 1 \\ 1 & 0}.
\end{equation}
This transformation cyclically permutes the three non-zero points of the phase space in a counter-clockwise sense: $(1,0) \rightarrow (1,1) \rightarrow (0,1) \rightarrow (1,0)$.
For $R$, and therefore for its powers, Eq.~(\ref{Amove}) does hold: there exists a unitary transformation $U_R$, unique up to a phase factor,
such that 
\begin{equation} \label{AmoveR}
U_R A_\alpha U^\dag_R = A_{R\alpha}.
\end{equation}
Specifically, 
\begin{equation}
U_R = \tfrac{1}{2}e^{i\pi/4}[ I - i(X+Y+Z)] = \frac{1}{\sqrt{2}}\mtx{cc}{1 & -i \\ 1 & i}.
\end{equation}
% CHANGE HERE
This transformation satisfies Eq.~(\ref{AmoveR}) because, when applied by conjugation to the Pauli matrices, it effects the permutation
\begin{equation} \label{XYZ}
X \rightarrow Y \rightarrow Z \rightarrow X.
\end{equation}
% CHANGE HERE
In fact, Eq.~(\ref{AmoveR}) requires this permutation, as one can see by writing the Pauli matrices as linear combinations of the $A$'s.
The other three symplectic matrices are of the form $R^s F$, where $F$ is the axis-interchanging transformation
\begin{equation}
F = \mtx{cc}{0 & 1 \\ 1 & 0}.
\end{equation}
In order to satisfy Eq.~(\ref{Amove}) with $L$ equal to $F$, we would need a unitary 
that effects the permutation
\begin{equation}
X \rightarrow Z \rightarrow X
\end{equation}
while leaving $Y$ entirely unchanged.  This transformation would be a reflection of the Bloch sphere through
a certain plane, which is not unitary.  Thus there is no correspondence between symplectic matrices and unitary operators
in the case of a single qubit, and a similar problem arises for any number of qubits \cite{Zhu,Raus1,Feldmann}.  

One way to deal with the problematic aspects of a system of qubits is to modify the phase space.  This approach has been taken in Ref.~\cite{Raussendorf2}, which employs an expanded phase
space with an overcomplete set of phase point operators, and in Ref.~\cite{Kocia}, in which the points of phase space are labeled by elements of a Grassmann algebra.  An earlier example of an expanded phase space for a system of qubits is the $2d \times 2d$ phase space proposed
by Leonhardt \cite{Leonhardt1, Leonhardt2}, for which the associated Wigner function has been shown to exhibit symplectic covariance \cite{Hashimoto}.
In the present paper we leave the $d \times d$ phase space unchanged but allow the possibility of reinterpreting its points and lines after a transformation.  We focus on a specific case, namely,
a two-qubit system, and work out what I expect is the simplest generalization of the property
(\ref{Wignermove}) that one can hope to apply to the phase space ${\mathbb F}_d^2$ when $d$ is a power of two.   

The idea can be illustrated in the single-qubit case by considering the above axis-interchanging transformation $F$.
Let us associate with this transformation the unitary operator
\begin{equation}
U_F = \frac{1}{\sqrt{2}}(Z - X) = \frac{1}{\sqrt{2}}\mtx{cc}{1 & -1 \\ -1 & -1}.
\end{equation}
Starting with a density matrix $\rho$, if we apply $U_F$ to $\rho$, the effect of $U_F$ is equivalent
to permuting the values of $W(\rho)$ in phase space according to $F$, {\em and} reinterpreting
these values in accordance with a different definition of the Wigner function.  Specifically,
one can show that
\begin{equation}  \label{reinterpret}
\widetilde{W}_\alpha(U_F \rho U_F^\dag) = W_{F^{-1}\alpha}(\rho),
\end{equation}
where the definition of $\widetilde{W}$ is closely analogous to that of $W$:
\begin{equation}
\widetilde{W}_\alpha(\rho) = \tfrac{1}{2} \hbox{Tr}(\tilde{A}_\alpha \rho),
\end{equation}
with
\begin{equation} \label{A00hat}
\tilde{A}_{(0,0)} =  \tfrac{1}{2}[I -X - Y -Z],
\end{equation}
and 
\begin{equation} \label{transhat}
\begin{split}
&\tilde{A}_{(1,0)} = X \tilde{A}_{(0,0)} X, \\ 
&\tilde{A}_{(1,1)} = Y \tilde{A}_{(0,0)} Y, \\
&\tilde{A}_{(0,1)} = Z \tilde{A}_{(0,0)} Z.
\end{split}
\end{equation}
That is, one obtains $\tilde{A}_\alpha$ for nonzero $\alpha$ just as before from the phase point operator associated with the 
origin, but the latter operator is now defined differently.
Thus it is still the case that we can express the effect of the unitary transformation $U_F$ by
letting $L$ tell us how to permute the values of the Wigner function, but we must reinterpret those
values with respect to a different operator basis, that is, a different set of phase point operators.  

For a single qubit, there is no need for more than the two operator bases we have seen here.  
For the case of two qubits, which is the main focus of this paper, we show below that we can
do something analogous using exactly
twelve operator bases, that is, twelve distinct sets of phase point operators.  Each such set can be generated by
specifying the phase point operator associated with the origin and then using standard displacement 
operators---generalizations of the $X$, $Y$, and $Z$ used above---to get the other $A$'s. 

Discrete Wigner functions have been applied with considerable success to problems in quantum computation.  
Typically---for example in Refs.~\cite{Raus1,Feldmann,Veitch1,Howard,Veitch2,Delfosse,Liu}---the phase space one uses for such 
applications, when the system's dimension is $d = p^n$ with $p$ a prime number, is the $2n$-dimensional vector space ${\mathbb Z}_p^{2n}$,
as proposed in Refs.~\cite{Wootters,Gross,Galetti}.  
This choice contrasts with the phase space we
use here, that is, the two-dimensional vector space ${\mathbb F}_d^2$ over the $d$-element field.  For many purposes 
the difference between the two phase spaces is inconsequential \cite{Gross}; however, it does make a difference
in the interpretation of Eq.~(\ref{Wignermove}).   As noted above, in this paper we take the matrix $L$ in Eq.~(\ref{Wignermove}) to be a $2 \times 2$ matrix with entries in ${\mathbb F}_d$, whereas
the higher-dimensional phase space naturally suggests making $L$ a
$2n \times 2n$ matrix with entries in ${\mathbb Z}_p$.  As a consequence, our focus will be on what has been called the ``restricted Clifford group'' rather than the
full Clifford group  \cite{Bengtsson,Zhu1,Zhu,Appleby}.  (We say more about this distinction in Section II below.)  Because of this difference, it is not immediately clear how the results we obtain in this paper might be relevant to the works
cited above on quantum computation.  On the other hand, other authors have used the ${\mathbb F}_d^2$ phase space to analyze quantum computation \cite{Galvao,Paz1,Paz2,vanDam}.
Moreover, there are other studies in which Eq.~(\ref{Wignermove}), with $L$ interpreted as we interpret it here, has played an important role \cite{Braasch1,Braasch2}, and the fact that that equation does not
apply to systems of qubits has limited the scope of those studies.  So there are reasons to look for a generalization of the equation in the ${\mathbb F}_d^2$ setting.  

It is a pleasure to offer this paper as a contribution to a special issue celebrating the birthday of Alexander Holevo.  In fact
one can identify a genuine historical chain connecting the work presented in this paper to some of Prof.~Holevo's work.  Much of his early work 
concerns the extraction of classical information from quantum systems \cite{Holevo1,Holevo2,Holevo3,Holevo4,Holevo5}, and these early results certainly motivated other researchers to
focus their attention on this problem.  Thinking about extracting 
classical information from quantum systems is precisely what led a number of authors to study
mutually unbiased bases, that is, orthonormal bases for a Hilbert space such that each vector in any one of the bases is an equal-magnitude superposition of the vectors in any
of the other bases.  Such bases are useful when one wants to extract classical information from a large collection of
identically prepared quantum systems \cite{Ivanovic,Fields,Petz}.  Finally, the discrete phase space we use in this paper is intimately
connected to mutually unbiased bases: as we describe in the following section, each striation of the phase space, that is,
each complete set of parallel lines, defines an orthogonal basis for the Hilbert space, and the bases associated with
different striations are mutually unbiased \cite{Gibbons,Bandyopadhyay,Pittenger}.  

Our construction in this paper follows the basic strategy of Ref.~\cite{Gibbons}, but we introduce, in Section III, a new element that makes the labeling
of the possible phase point operators more systematic.  
First, though, we describe in Section II the discrete phase space and the displacement 
operators for a system of two qubits, and we show how the action of each linear symplectic transformation $L$
on the points of this phase space can be mimicked by the action of a corresponding unitary operator $U_L$ on the displacement operators.  Section III shows how to construct,
non-uniquely, a discrete Wigner function
satisfying two requirements: (i) the Wigner function must be covariant under displacements, and (ii) the sums of the Wigner function over the lines in phase space having a particular slope must equal the probabilities of the outcomes of the orthogonal measurement associated with that slope.   It is at this point that we introduce our
systematic labeling scheme, and we single out
one particular definition of the Wigner function to call ``standard.''  To keep track of the effect
of the unitaries $U_L$ on the phase point operators, we introduce in Section IV the notion of the ``index'' of a phase point operator.  
Shifting to a different definition of the Wigner function, that is, shifting to a different interpretation of the Wigner-function values, is a matter of changing the index of the phase point
operator associated with the origin of phase space.  We describe such reinterpretations in Section V and give a couple of examples.  Finally, we summarize and discuss the results
in Section VI.  
  
Though in this paper we focus on the case of just two qubits, there is no reason to doubt that many of our results
can be extended to an arbitrary number of qubits.  Our construction relies heavily on the existence of a symplectic linear transformation that 
cycles through all the possible slopes of phase-space lines---we think of it as a rotation---and it is known that such a transformation exists for the $d \times d$ finite-field phase space
whenever $d$ is a power of two \cite{Chau,Sussman}.  So it is reasonable to hope such an extension can be found.
In the meantime, even our results for a pair of qubits give us a novel way of representing operations on this simple but conceptually interesting quantum system.

\section{A two-qubit phase space}

Again, the phase space we consider here is a two-dimensional vector space over a finite field.  For a system of 
two qubits, the relevant field is the four-element field, ${\mathbb F}_4$.
Let the elements of this field be called $\{0,1,\omega,\bar{\omega}\}$.  Addition and multiplication in this field are
given by the following tables.

\bigskip

\begin{centering}

\begin{tabular}{c!{\vrule width 1pt}c|c|c|c|}
$+$ & 0 & 1 & $\omega$ & $\bar{\omega}$ \\
\noalign{\hrule height 1pt}
0 & 0 & 1 & $\omega$ & $\bar{\omega}$ \\
\hline 
1 & 1 & 0 & $\bar{\omega}$ & ${\omega}$ \\
\hline 
$\omega$ & $\omega$ & $\bar{\omega}$ & 0 & 1 \\
\hline
$\bar{\omega}$ & $\bar{\omega}$ & ${\omega}$ & 1 & 0 \\
\hline
\end{tabular}
\hspace{2cm}
\begin{tabular}{c!{\vrule width 1pt}c|c|c|c|}
$\times$ & 0 & 1 & $\omega$ & $\bar{\omega}$ \\
\noalign{\hrule height 1pt}
0 & 0 & 0 & 0 & 0 \\
\hline 
1 & 0 & 1 & ${\omega}$ & $\bar{\omega}$ \\
\hline 
$\omega$ & 0 & ${\omega}$ & $\bar{\omega}$ & 1 \\
\hline
$\bar{\omega}$ & 0 & $\bar{\omega}$ & 1 & $\omega$ \\
\hline
\end{tabular}

\end{centering}

\bigskip

\noindent As in the Introduction, a point $\alpha$ in the phase space has
horizontal and vertical coordinates $\alpha_q$ and $\alpha_p$; these
coordinates now take values in ${\mathbb F}_4$.  So the two-qubit phase space
can be pictured as a $4 \times 4$ grid.  

The field arithmetic allows us to identify lines and parallel lines in the phase space: 
a line consists of points $\alpha$ satisfying an equation of the form
\begin{equation}
a \alpha_q + b \alpha_p = c,
\end{equation}
where $a$, $b$, and $c$ are in ${\mathbb F}_4$ with $a$ and $b$ not both zero.  Two lines are parallel if they can be represented
by equations of this form that differ only in the value of $c$.  A complete set of parallel lines is called a striation---in this $4 \times 4$ phase
space there are exactly five striations.  It will also be useful to refer to
the {\em rays} in phase space, which are the lines running through the origin (so that $c$ equals zero).  

In order to distinguish our two qubits, we choose, for each of the two phase-space axes, a {\em basis} for the field, that is,
a pair of field elements $(e_1, e_2)$ such that any element $x$ of the field can
be written as 
\begin{equation}
x = x_1 e_1 + x_2 e_2,
\end{equation}
$x_1$ and $x_2$ being elements of ${\mathbb F}_2 = \{0,1\}$.  The coefficients $x_1$ and $x_2$ will be 
associated with the first and second qubits, respectively, in a sense to be explained in the following paragraph.  In order to guarantee the connection with mutually unbiased bases that we rely on in our definition of the Wigner function,
we need to choose the bases we assign to the two axes to be the {\em duals} of each other, at least up to a multiplicative constant \cite{Gibbons, Bengtsson, Appleby}.  
For the rest of this paper we adopt the specific basis $(e_1,e_2) = (\bar{\omega}, {\omega})$ for each axis.  This basis happens to be self-dual, thus giving us a certain
symmetry between the two axes.  

A {\em displacement} of the phase space by the vector $\beta$ is a mapping that takes
each point $\alpha$ of phase space to the point $\alpha + \beta$.
With each such displacement we associate a unitary transformation as follows.
Let $\beta = (\beta_q, \beta_p)$ be the displacement vector.  Here $\beta_q$ and $\beta_p$
are elements of ${\mathbb F}_4$, and they can be expanded in the 
basis $(\bar{\omega}, {\omega})$:
\begin{equation}
\beta_q = \beta_{q1}\bar{\omega} + \beta_{q2}{\omega} , \hspace{1cm}  \beta_p = \beta_{p1}\bar{\omega} + \beta_{p2}{\omega}.
\end{equation}
We define the unitary displacement operator $D_\beta$
to be
\begin{equation}  \label{Dpropto}
D_\beta = e^{i \phi} X^{\beta_{q1}} Z^{\beta_{p1}} \otimes X^{\beta_{q2}}Z^{\beta_{p2}}
\end{equation}
with a phase $\phi$ that may depend on $\beta$.  
To fix this phase, we choose each displacement operator
to be of the form $D_\beta = \sigma_j \otimes \sigma_k$, where the $\sigma$'s are the 
matrices $\{ I, X, Y, Z\}$.  Thus, when Eq.~(\ref{Dpropto}) calls on us to use the product $XZ$ for, say, the
first qubit, we write $Y$ in place of $XZ$.  This choice makes each displacement operator 
Hermitian in addition to being unitary.  Note that the displacement operators satisfy the composition rule
\begin{equation} \label{adddisplacements}
D_\alpha D_\beta \propto D_{\alpha + \beta},
\end{equation}
where the proportionality sign indicates that there may be a nontrivial phase factor relating $D_\alpha D_\beta$ to $D_{\alpha + \beta}$.
In fact the phase factor will always be a power of $i$ \cite{Calderbank}.
The following table shows the discrete phase space, with the axes labeled by the values of $\beta_q$ and
$\beta_p$ and with each displacement operator $D_\beta$ indicated at the location of the point $\beta$.  

\bigskip

\begin{centering}

\hspace{-2.4cm}
\begin{tabular}{cc}

\hspace{-0.4cm} $\beta_p$ \hspace{-2.2cm} &

\begingroup

\hspace{0.8cm}
\setlength{\tabcolsep}{3pt} % Default value: 6pt
\renewcommand{\arraystretch}{1.5} % Default value: 1
\begin{tabular}{cc}
\begin{tabular}{c}
$\bar{\omega}$ \\
$\omega$ \\
1 \\
0 \\
\end{tabular} &
\begin{tabular}{|c|c|c|c|}
\hline
 $Z\otimes I$ & $Y\otimes X$ & $Z\otimes X$ & $Y\otimes I$ \\
\hline
% CHANGE HERE
$I\otimes Z$ & $X\otimes Y$ & $I\otimes Y$ & $X\otimes Z$ \\
\hline
$Z\otimes Z$ & $Y\otimes Y$ & $Z\otimes Y$ & $Y\otimes Z$ \\
\hline
$I\otimes I$ & $X\otimes X$ & $I\otimes X$ & $X\otimes I$ \\
\hline
\end{tabular} \\
 &\setlength{\tabcolsep}{14.7pt}  \begin{tabular}{cccc} 
 \hspace{-0.8mm}0 & 1 & $\omega$ & $\bar{\omega}$ 
 \end{tabular}
 \end{tabular}

\endgroup \\

 & \hspace{1.7cm} $\beta_q$ 
 
 \end{tabular}
 
 \end{centering}

\bigskip

\noindent Throughout this paper, we consistently label the phase-space axes as they are labeled in this table, with the field values $0, 1, \omega, \bar{\omega}$ always 
appearing in that order.  

Now consider any ray.  Referring to the above 
table, we see that each ray is associated with a set of four displacement operators,
one of them being the identity.  One can check that these four operators commute
with each other.  For example, the ray with slope $\omega$ consists of the points $(0,0), (1,\omega), (\omega, \bar{\omega})$, and $(\bar{\omega}, 1)$, and the associated 
displacement operators are $I\otimes I, X \otimes Y, Z\otimes X$, and $Y\otimes Z$.  Since the operators commute,
there exists an orthonormal basis of simultaneous eigenvectors of these operators.
In fact, for each of the five rays, the basis is unique up to phase factors.  We associate this basis with the given ray as well as with the striation that includes that ray.  
One can show that the five bases generated in this way are mutually unbiased---this result in fact generalizes to all prime-power values of the Hilbert-space dimension \cite{Gibbons,Bandyopadhyay,Pittenger}.
Moreover, for {\em any} dimension $d$, the number of
mutually unbiased bases cannot be greater than $d+1$ \cite{Ivanovic}.  So the five bases associated with the five striations constitute a complete set
of mutually unbiased bases for the four-dimensional Hilbert space.  
Again, these mutually unbiased bases play a central role in defining a Wigner function
for two qubits, as we discuss in the following section.  

For now, we turn our attention to symplectic linear transformations, represented by unit-determinant $2 \times 2$ matrices with entries in ${\mathbb F}_4$.  
For each such matrix $L$, there exists a unitary transformation $U$ such that
\begin{equation} \label{ULonD}
U D_\alpha U^\dag \propto D_{L\alpha},
\end{equation}
where the possible phase factor indicated by the proportionality sign is now $\pm 1$ \cite{Calderbank}.  
In fact, for each $L$ there are exactly 16 such unitary transformations $U$ differing from each other by more than just a phase factor.  If $U$ and $U'$ are two such unitaries for 
a given $L$, then there exists a displacement operator $D_\beta$ such that $U' \propto D_\beta U$  \cite{Calderbank,Zhu}.  The number 16 thus comes from the fact that there
are 16 displacement operators.  

When the Hilbert-space dimension $d$ is odd, one can write the analog of Eq.~(\ref{ULonD}) with an equals sign rather than a proportionality sign, and the unitary corresponding to $L$
is then unique up to a phase factor \cite{Appleby}. 
% It is in this setting that one obtains a unique Wigner function satisfying Eq.~(\ref{Wignermove}) \cite{Gross,Zhu3}.  
Unfortunately, within our 
two-qubit framework the equals sign
is not possible.  To see this, it is enough to consider again the axis-interchanging transformation $\mtx{cc}{0 & 1 \\ 1 & 0}$.  If there were an equals sign in Eq.~(\ref{ULonD}), the unitary $U$ associated with this matrix would have
to satisfy 
\begin{equation}
\begin{split}
&U(X \otimes I)U^\dag = Z \otimes I, \\
&U(Z \otimes I)U^\dag = X \otimes I, \\
&U(Y \otimes I)U^\dag = Y \otimes I.
\end{split}
\end{equation}
But by multiplying the first two of these equations together, we get \hbox{$U(Y \otimes I)U^\dag = -Y \otimes I$}, which contradicts the third; moreover, this argument would be unaffected by any 
additional phase factors we might attach to the displacement operators.  So for us there is not a unique choice of unitary for each $L$.  (One can achieve equality in Eq.~(\ref{ULonD}) for a system of qubits by letting $L$ be a $2 \times 2$ matrix with entries in ${\mathbb Z}_{2d}$, which for the two-qubit case would be ${\mathbb Z}_8$ \cite{ApplebyJMP}.  But this choice would entail a rather different framework from the one in which we are working.)

For definiteness, though, we now choose one particular unitary matrix satisfying Eq.~(\ref{ULonD}) for each symplectic $L$, and we distinguish this unitary with the symbol $U_L$.  
We make our choices in such a way that the $U_L$'s also satisfy the equation
\begin{equation} \label{rep}
U_{L_1L_2} \propto U_{L_1}U_{L_2}.
\end{equation}
We begin by 
identifying a unitary matrix for each of the four symplectic matrices $H_x$ of the form $H_x = \mtx{cc}{1 & 0 \\ x & 1}$ with $x \in {\mathbb F}_4$.  
We can think of these symplectic transformations as vertical shears, as they shift the points within each vertical line while leaving the line
itself unchanged.   
\begin{align}  \label{UL}
\begin{split}
H_0 = I = \mtx{cc}{1 & 0 \\ 0 & 1}\hspace{1cm} & \setlength\arraycolsep{3.4pt}  U_{H_0} =  \mtx{cccc}{1 & 0 & 0 & 0 \\ 0 & 1 & 0 & 0 \\ 0 & 0 & 1 & 0 \\ 0 & 0 & 0 & 1 } ,   \\
\fixTABwidth{T} 
H_1 = \mtx{cc}{1 & 0 \\ 1 & 1}\hspace{1cm} & \fixTABwidth{T}  U_{H_1} = \parenMatrixstack{0 & 0 & 0 & -i \\ 0 & 0 & 1 & 0 \\ 0 & 1 & 0 & 0 \\ i & 0 & 0 & 0} ,  \\
\fixTABwidth{T}
H_\omega = \mtx{cc}{1 & 0 \\ \omega & 1}\hspace{1cm} & \fixTABwidth{T}  U_{H_\omega} =  \parenMatrixstack{ 0 & -i & 0 & 0 \\ i & 0 & 0 & 0 \\ 0 & 0 & 0 & 1 \\ 0 & 0 & 1 & 0 } ,   \\
\fixTABwidth{T}
H_{\bar{\omega}} = \mtx{cc}{1 & 0 \\ \bar{\omega} & 1 }\hspace{1cm} & \fixTABwidth{T}  U_{H_{\bar{\omega}}} = \parenMatrixstack{ 0 & 0 & -i & 0 \\ 0 & 0 & 0 & 1 \\ i & 0 & 0 & 0 \\ 0 & 1 & 0 & 0 } .
\end{split}
\end{align}
We also choose a unitary operator to associate with the following symplectic matrix $R$, which we think of as a rotation, since it cycles through all five 
striations.
\begin{equation}  \label{RUR}
R = \mtx{cc}{\bar{\omega} & 1 \\ 1 & 0} \hspace{1cm} \fixTABwidth{T} \setstacktabbedgap{1.5mm} U_R = \frac{1}{2}\parenMatrixstack{i & 1 & i & -1 \\ i & -1 & i & 1 \\ i & 1 & -i & 1 \\ i & -1 & -i & -1}.
\end{equation}
One can verify that the following sixty matrices are distinct and therefore represent all sixty of the $2 \times 2$ symplectic matrices with entries in ${\mathbb F}_4$.  
\begin{equation}  \label{RHR}
\begin{split}
&H_0 R^s, \;\; s=0, \ldots, 4, \\
&H_{\bar{\omega}}R^s,  \;\; s=0, \ldots, 4, \\
&R^r H_1 R^s,  \;\; r,s=0, \ldots, 4, \\
&R^r H_\omega R^s,   \;\; r,s=0, \ldots, 4. \\
\end{split}
\end{equation}
($H_{\bar{\omega}}$ has the special property that $RH_{\bar{\omega}}R = H_{\bar{\omega}}$.  So it would be redundant to multiply $H_{\bar{\omega}}$ by $R$ on both sides.)
For each such matrix $L$, we define its associated unitary $U_L$ to be the operator obtained by replacing each $H$ with the appropriate $U_H$ and replacing each
$R$ with $U_R$.  Thus, for example, for the matrix
\begin{equation}  \label{exampleL}
G = R^2 H_1 R = \mtx{cc}{\bar{\omega} & 0 \\ 0 & \omega },
\end{equation}
the associated unitary is 
\begin{equation}  \label{exampleUL}
\fixTABwidth{T}
U_G = U_R^2 U_{H_1} U_R = \parenMatrixstack{-1& 0 & 0 & 0 \\ 0 & 0 & 1 & 0 \\ 0 & 0 & 0 & -1 \\ 0 & 1 & 0 & 0 }.
\end{equation}
% CHANGE HERE
That the $U_L$'s defined in this way satisfy Eq.~(\ref{rep}) is shown in the Appendix.

If we are given a symplectic matrix $L$, then in order to apply the above definition of $U_L$, it would evidently be helpful to know how to express $L$ in one of the four forms seen
in Eq.~(\ref{RHR}).  One way to do this is as follows.  First, to find the value of $x$ in $L = R^r H_x R^s$, we can make use of the fact that $RH_{\bar{\omega}}R=H_{\bar{\omega}}$, from which one can
show that
\begin{equation}
\hbox{Tr}(L^T H_{\bar{\omega}}L H_{\bar{\omega}}) = x^2.
\end{equation}
(The $H_{\bar{\omega}}$'s in this equation essentially remove all the $R$'s that appear in $L$, leaving only $H_x$.)  Then, if $x$ turns out to be $0$ or $\bar{\omega}$, we can find the value of
$s$ in $L=H_x R^s$ by computing the vector $\mu=L \mtx{c}{1 \\ 0}$.  The ratio $\mu_p/\mu_q$, which takes five possible values, including ``infinity'' when $\mu_q$ is zero, turns out to be perfectly correlated with the value of $s$:

\bigskip

\begin{centering}

\begin{tabular}{c|x{4.7mm}x{4.7mm}x{4.7mm}x{4.7mm}c}
$x+ \mu_p/\mu_q $ & $\omega$ & $\bar{\omega}$ & $0$ & $1$ & $\,\infty\,$ \\
\hline
$s$ & 1 & 3 & 0 & 2 & 4 \\
\end{tabular}  \hspace{0.7cm} ($x = 0$ or $\bar{\omega}$).

\end{centering}

\bigskip

\noindent (Here $x + \infty$ is understood to be $\infty$.) If $x$ comes out to be $1$ or
$\omega$, we can isolate the integer $s$ in $L=R^r H_x R^s$ by computing the vector
\begin{equation}
\nu =\left[ L^T H_{\bar{\omega}} L  \mtx{c}{1 \\ 0}\right] -  \mtx{c}{1 \\ x},
\end{equation}
whose ``slope'' $\nu_p/\nu_q$ is likewise perfectly correlated with the value of $s$:

\bigskip

\begin{centering}

\begin{tabular}{c|x{4.7mm}x{4.7mm}x{4.7mm}x{4.7mm}c}
$\nu_p/\nu_q $ & $\omega$ & $\bar{\omega}$ & $0$ & $1$ & $\,\infty\,$ \\
\hline
$s$ & 0 & 1 & 2 & 3 & 4 \\
\end{tabular}  \hspace{0.7cm} ($x = 1$ or ${\omega}$).

\end{centering}

\bigskip

 \noindent And we can similarly get the value of $r$ from the slope of 
\begin{equation}
\tau = \left[ L H_{\bar{\omega}} L^T  \mtx{c}{1 \\ 0} \right] -  \mtx{c}{1 \\ x},
\end{equation}
in accordance with the following table:

\bigskip

\begin{centering}

\begin{tabular}{c|x{4.7mm}x{4.7mm}x{4.7mm}x{4.7mm}c}
$\tau_p/\tau_q $ & $\omega$ & $\bar{\omega}$ & $0$ & $1$ & $\,\infty\,$ \\
\hline
$r$ & 1 & 2 & 3 & 4 & 0 \\
\end{tabular}  \hspace{0.7cm} ($x = 1$ or ${\omega}$).

\end{centering}

\bigskip

\noindent Once we know the values of $x$, $r$, and $s$---here $r$ is taken to be zero if $x=0$ or $x=\bar{\omega}$---we identify the unitary $U_L$ as
\begin{equation}
U_L = U_R^r U_{H_x} U_R^s.
\end{equation}

The set of unitary operators of the form $D_\beta U_L$, modulo phase factors, constitutes 
the restricted Clifford group mentioned in the Introduction \cite{Bengtsson,Zhu1,Zhu,Appleby}.  It is a subgroup of the full Clifford group, differing from this 
latter group in that Eq.~(\ref{ULonD}) respects the finite-field arithmetic---in our case the arithmetic of ${\mathbb F}_4$.  For example,
% CHANGE HERE
the controlled-not operation permutes the displacement operators and is thus a Clifford operation, but the permutation is not linear
in ${\mathbb F}_4^2$.
So the controlled-not operation is not an element of our restricted Clifford group.

\section{A two-qubit Wigner function}

As we have noted in the preceding section, each striation of the phase space is associated with an orthogonal basis of the Hilbert space,
and the five bases obtained in this way are mutually unbiased.  The Wigner functions of Ref.~\cite{Gibbons} are defined specifically so that
the sums of $W_\alpha$ over the lines of any striation yield the probabilities of the outcomes of the measurement defined by the associated
Hilbert-space basis.  An analogous property holds for the original, continuous Wigner function \cite{Wootters,Bertrand}.  Another key idea of Ref.~\cite{Gibbons} is
that the Wigner function should be covariant under displacements.  That is, 
\begin{equation}  \label{dispcov}
W_\alpha(D_\beta \rho D_\beta^\dag) = W_{\alpha - \beta}(\rho).
\end{equation}
(The dagger in this equation is unnecessary, since our $D_\beta$'s are Hermitian.  We include it so that the equation
would still hold if one were to adopt a different phase convention for the $D_\beta$'s.)
These requirements still allow many distinct definitions of the Wigner function.  The following algorithm will generate 
every Wigner-function definition compatible with these requirements \cite{Gibbons}. 
\begin{enumerate}
\item For each ray $n$, choose arbitrarily a vector from the orthogonal basis associated with the ray.  (In Fig. 2 below, we label the rays and striations with
the values $n=0, \ldots, 4$, but we do not need the details of that labeling scheme here.)  Let the vectors that have 
been chosen be called $|v^{(n)}\rangle$.
\item Construct the operator
\begin{equation}  \label{originoperator}
A_0 = \sum_n |v^{(n)}\rangle \langle v^{(n)}| - I,
\end{equation}
where the sum is over all the rays.
Take this operator to be the phase point operator associated with the origin of phase space. 
\item Define the other phase point operators by the following equation:
\begin{equation}  \label{displaceAA}
A_\alpha = D_\alpha A_0 D^\dag_\alpha.
\end{equation}
\item The Wigner function $W_\alpha(\rho)$ is then defined by
\begin{equation}
W_\alpha(\rho) = \tfrac{1}{4}\hbox{Tr}(A_\alpha \rho).
\end{equation}
\end{enumerate}
Now,
for each of the five rays, there are four possible vectors to choose in step 1, and after these choices are made the definition 
is fully determined.  So there are $4^5 = 1024$ possible definitions of the two-qubit Wigner function.

It will be helpful to have a way of labeling these possible definitions.  Alternatively, we can say we are labeling the
1024 possible choices of the ``origin operator'' $A_0$.  And it will be helpful to make this labeling very systematic.

\begin{figure} 
\centering
\includegraphics[scale=0.76]{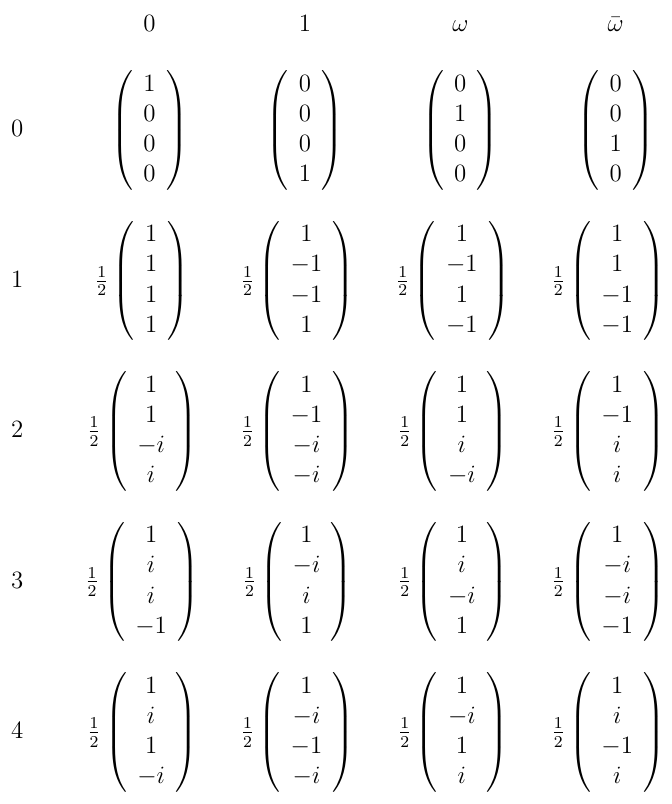}
\caption{Labeling scheme for the vectors of the five mutually unbiased bases.  The integer $n$ at the left labels the bases, and the field element $k_n$ 
at the top labels the vectors of each basis.  (The vectors shown here sometimes differ from $|b_k^{(n)}\rangle$ by an overall phase factor.)}   \label{vectortable} 
\end{figure}

To this end, we return to the unitary operator $U_R$ of Eq.~(\ref{RUR}).  
We use $U_R$ to construct a labeling scheme for the vectors of the five special mutually unbiased bases.
We start with the standard basis, which we refer to as the zeroth basis.  Let us call the vectors in this basis
$(|b^{(0)}_0\rangle, |b^{(0)}_1\rangle, |b^{(0)}_\omega\rangle, |b^{(0)}_{\bar{\omega}}\rangle)$, where $|b^{(0)}_0\rangle$ is the column vector 
with components $(1, 0, 0, 0)$ and
$|b^{(0)}_k\rangle = D_{(k,0)}|b^{(0)}_0\rangle$.  We then obtain the other bases, and their labels, by 
repeatedly applying $U_R$.  For $n=0, \ldots, 4$, which labels the bases, and for $k=0,1,\omega,\bar{\omega}$, which labels the vectors
within each basis, we define
\begin{equation}
|b^{(n)}_k\rangle  = U_R^n|b^{(0)}_k\rangle = U_R^n D_{(k,0)} |b^{(0)}_0\rangle.
\end{equation}
It is worth writing down explicitly the vectors of the five bases, along with their labeling.  We do this in Fig.~\ref{vectortable}.

Again, the origin operator $A_0$ is obtained by choosing one vector from each of these five bases, and it can now be
expressed as 
\begin{equation}  \label{A0ks}
A_0 = \sum_{n=0}^4 |b^{(n)}_{k_n}\rangle\langle b^{(n)}_{k_n} | - I.
\end{equation}
Let us choose a particular definition of the origin operator $A_0$, and hence of the Wigner function, and distinguish this definition with the adjective ``standard'':  it is
the one for which $A_0$ has $k_0 = k_1 = k_2 = k_3 = k_4 = 0$.  Notationally,
we distinguish the standard Wigner function, and its associated phase point operators, with a boldfaced ${\bf W}$ and ${\bf A}$.  

%Every definition of the Wigner function---that is, every choice of the origin operator $A_0$---defines a correspondence between
%the lines of phase space and the vectors of the mutually unbiased bases.  It is the correspondence expressed in the symbol $|v_\lambda\rangle$ of Eq.~(\ref{altstep3}).
%Or one could put it in the following terms: the basis vector $|v_\lambda\rangle$ corresponds to the line $\lambda$ if summing the Wigner function over $\lambda$ yields the probability of obtaining the measurement outcome associated with $|v_\lambda\rangle$.  

\begin{figure} 
\centering
\includegraphics[scale=0.65]{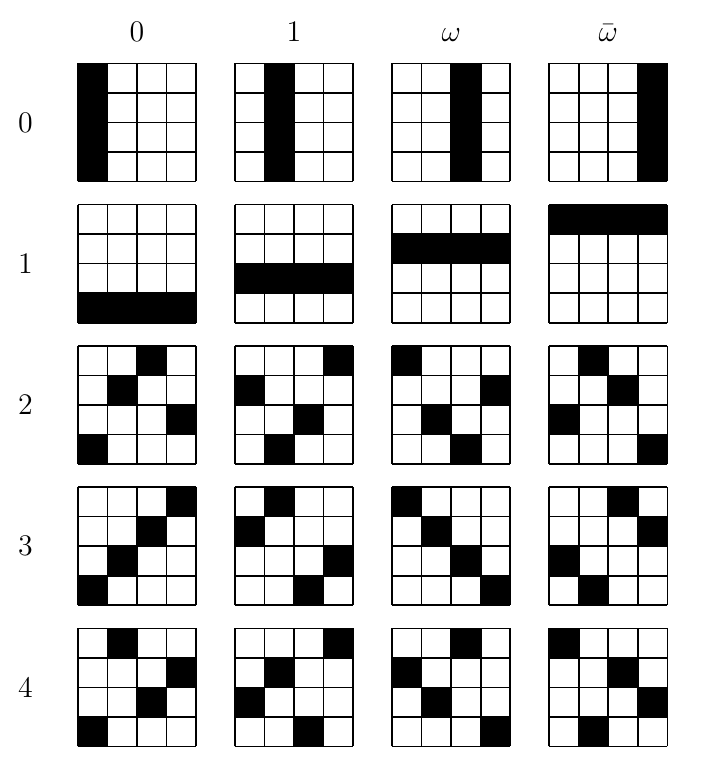}
\caption{Striations obtained by repeatedly applying the linear transformation $R$, starting with the vertical striation.   The integer $n$ at the left labels the
striations, and the field element $k_n$ at the top labels the lines within each striation.   
%Our {\em standard} definition of the 
%Wigner function associates each line in this figure with the Hilbert-space vector located at the same place in Fig.~\ref{vectortable}. 
}  \label{striations} 
\end{figure}

%Let us make this correspondence explicit in the case of the standard Wigner function.  To do this, we first 
We would also like to label the lines of phase space.  We do so in a way that parallels our labeling of the basis vectors.  
Fig.~\ref{striations}, which shows all the lines, was constructed as follows: at the top are the vertical lines, which we label $\lambda^{(0)}_k$ with $k \in {\mathbb F}_4$.
Each successive entry of the figure is obtained from the entry above it by applying our ``rotation'' matrix $R$.  
Thus we define $\lambda^{(n)}_k$ to be $R^n \lambda^{(0)}_k$.  (Applying $R$ to a line means applying it to each point in the line.)  
%Our standard Wigner function ${\bf W}$
%associates the Hilbert-space vector $|b^{(n)}_k\rangle$ with the line $\lambda^{(n)}_k$.  That is,
%\begin{equation}
%\sum_{\alpha \in \lambda^{(n)}_k}{\bf W}_\alpha(\rho) = \langle b^{(n)}_k |\rho | b^{(n)}_k\rangle.
%\end{equation}

\bigskip

\section{Behavior of the ``index'' of a phase point operator or set of lines}

Regardless of which choice we have made for $A_0$---whether ``standard'' or not---any two-qubit phase point operator $A$ can be expressed as
in Eq.~(\ref{A0ks}):
\begin{equation}  \label{Afrombs}
A = \sum_{n=0}^4 |b^{(n)}_{k_n}\rangle\langle b^{(n)}_{k_n} | - I.
\end{equation}
We define the 
{\em index} of the operator $A$, denoted ${{\mathcal I}}(A)$, to be the five-component column vector whose components are the $k_n$'s in the above equation.  

The concept of an index can also be applied to a collection of {\em lines}.  Let $\Lambda$ be any ordered set of lines of the form
$\Lambda = (\lambda^{(0)}_{k_0}, \lambda^{(1)}_{k_1}, \lambda^{(2)}_{k_2}, \lambda^{(3)}_{k_3}, \lambda^{(4)}_{k_4})$; that is, $\Lambda$ picks out one line
from each striation.  We say that
the index of $\Lambda$ is the five-component column vector whose entries are the $k_n$'s.  Note, in particular, that any {\em point} $\alpha$ of phase space
defines such a set $\Lambda$: it picks out the line in each striation on which $\alpha$ lies.  So we can also speak of the index of a point. 
For example, by consulting Fig.~\ref{striations}, one can see that if $\alpha$ is the point $(1,\omega)$, its index is
\begin{equation}
{\mathcal I}(\alpha) = \mtx{c}{1 \\ \omega \\ 0 \\ \omega \\ 1}.
\end{equation}

We now ask how the index, of a collection of lines or of a phase point operator, is affected by (i) a displacement and (ii) a symplectic linear transformation.  

\subsection{The effect of a displacement}

We begin with the case of a displacement, and for now we focus on a collection $\Lambda$ of lines.  Let $\Delta_\beta$ represent 
a displacement by $\beta$.  ($\Delta_\beta$ acts on phase-space points, whereas $D_\beta$ acts on Hilbert-space vectors.)  
How does the index of $\Lambda$ change under a displacement?  A displacement will move each line to a line in the same
striation; so the $k$-value of the line might change but the $n$-value will not.  We have
\begin{equation}
\begin{split}
\Delta_{\beta} \lambda^{(n)}_k &= \Delta_{\beta} R^n \lambda^{(0)}_k \\
&= R^{n}(R^{-n}\Delta_{\beta} R^n) \lambda^{(0)}_k \\
&= R^n \Delta_{R^{-n}\beta} \lambda^{(0)}_k .
\end{split}
\end{equation}
Now, the $k$-value of a vertical line is not changed by a vertical displacement; so we need only the horizontal
component of $R^{-n}\beta$:
\begin{equation}
\Delta_{\beta} \lambda^{(n)}_k = R^n   \lambda^{(0)}_{k + (R^{-n}\beta)_q} =  \lambda^{(n)}_{k + (R^{-n}\beta)_q}.
\end{equation}
For each value of $n$, the expression $(R^{-n}\beta)_q$ is linear in $\beta$, so it must be of the form
\begin{equation}
(R^{-n}\beta)_q =  \beta_q Q_n+\beta_p P_n,
\end{equation}
where $Q$ and $P$ are five-component column vectors with entries in ${\mathbb F}_4$.  In fact one finds that
\begin{equation} \label{hv}
Q = \mtx{c}{1 \\ 0 \\ 1 \\ \bar{\omega} \\ \bar{\omega} }, \hspace{1cm} P = \mtx{c}{0 \\ 1 \\ \bar{\omega} \\ \bar{\omega} \\ 1 }.
\end{equation}
We thus have
\begin{equation}  \label{displacelambda}
{\mathcal I}(\Delta_\beta \Lambda) = {\mathcal I}(\Lambda) + \beta_q Q + \beta_p P.
\end{equation}
%As a special case, for a phase space point $\alpha$, we can write
%\begin{equation}
%{\mathcal I}(\Delta_\beta \alpha) = {\mathcal I}(\alpha) + \beta_q Q + \beta_p P.
%\end{equation}

We can make essentially the same argument to find the effect of a unitary displacement operator $D_\beta$
on the index of a phase point operator, and we obtain essentially the same answer.  Specifically, we can write
\begin{equation}
\begin{split}
D_\beta |b^{(n)}_k\rangle &= D_\beta U_R^n |b^{(0)}_k\rangle \\
&= U_R^n (U^{-n}_R D_\beta U^n_R) |b^{(0)}_k\rangle \\
&\propto U_R^n D_{R^{-n}\beta} |b^{(0)}_k\rangle . \\
\end{split}
\end{equation}
But $|b^{(0)}_k\rangle$ is an eigenstate of any unitary displacement operator associated with a vertical displacement.
So we need to consider only the horizontal component of $R^{-n}\beta$:
\begin{equation}  \label{xxx}
\begin{split}
D_\beta |b^{(n)}_k\rangle \propto U_R^n |b^{(0)}_{k + (R^{-n}\beta)_q}\rangle &= |b^{(n)}_{k + (R^{-n}\beta)_q}\rangle \\
&= |b^{(n)}_{k + \beta_q Q_n + \beta_p P_n}\rangle .
\end{split}
\end{equation}
Since
each of our phase point operators is built out of the vectors $|b^{(n)}_k\rangle$ in accordance with 
Eq.~(\ref{Afrombs}), Eq.~(\ref{xxx}) leads to the equation
\begin{equation} \label{displaceA}
{\mathcal I}(D_\beta A_\alpha D^\dag_\beta) = {\mathcal I}(A_\alpha) + \beta_q Q + \beta_p P.
\end{equation}
We see, then, that the change in the index of either a set of lines or a phase point operator upon displacement by $\beta$ is fairly straightforward:
we simply add $\beta_q Q + \beta_p P$.

\subsection{The effect of a symplectic linear transformation}

We now turn our attention to the case of a symplectic linear transformation $L$.  We begin by considering the effect of 
$L$ on the index of a set of lines $\Lambda$.  Unlike a displacement, a linear transformation can take a striation to 
a different striation.  But at least it will not split up a striation: if two lines are parallel, their transformed images will also
be parallel.  For a given linear transformation $L$, the $n$th striation (where $n=0, \ldots, 4$) will be sent by $L$ 
to some striation whose label we call $m$.  (The value of $m$ depends on $n$, but we leave this dependence implicit.)
The transformation $L$ will send rays to rays, so 
\begin{equation}
L \lambda^{(n)}_0 = \lambda^{(m)}_0.
\end{equation}
To figure out the effect of $L$ on the labels of the other lines, we use the fact that
\begin{equation}
\Delta_\beta \lambda^{(n)}_0 = \lambda^{(n)}_{\beta_q Q_n + \beta_p P_n}.
\end{equation}
Let us now apply $L$ to the right-hand side of this equation.
\begin{equation}  \label{bigone}
\begin{split}
L \lambda^{(n)}_{\beta_q Q_n + \beta_p P_n} &= L\Delta_\beta \lambda^{(n)}_0 \\
&= (L\Delta_\beta L^{-1}) L \lambda^{(n)}_0 \\
&= \Delta_{L\beta} L\lambda^{(n)}_0 \\
&=\Delta_{L\beta} \lambda^{(m)}_0 \\
&=\lambda^{(m)}_{(L\beta)_q Q_m + (L\beta)_p P_m}.
\end{split}
\end{equation}
For each value of $n$, any value of $k$ can be written as $\beta_q Q_n + \beta_p P_n$ for at least one value of $\beta$.  
So the above equation is sufficient to tell us what $L$ does to each of the lines.  
%In fact, the equation tells us how the $m$th component of the index of $L\Lambda$ is determined by the
%$n$th component of the index of $\Lambda$:
%\begin{equation}
%{\mathcal I}(L\Lambda)_m = \frac{(L \beta)_q Q_m + (L \beta)_p P_m}{\beta_q Q_n + \beta_p P_n} {\mathcal I}(\Lambda)_n
%\end{equation}
Moreover, to multiply $k$ by any factor, we can multiply $\beta$
by the same factor, and the equation tells us that the subscript of $\lambda^{(m)}$ is also multiplied by that same factor.  
So for a set of lines $\Lambda$, the $m$th component of the index of $L\Lambda$ must be simply proportional to the $n$th component of the index of $\Lambda$ (where
the proportionality constant is an element of ${\mathbb F}_4$).
According to Eq.~(\ref{bigone}), the relation must be expressible as
\begin{equation}
{\mathcal I}(L\Lambda)_m = \left( \frac{(L \beta)_q Q_m + (L \beta)_p P_m}{\beta_q Q_n + \beta_p P_n}\right) {\mathcal I}(\Lambda)_n
\end{equation}
for any value of $\beta$ that makes the denominator nonzero.  (The value of the fraction is the same for any such value of $\beta$.)
We can write this relation as
\begin{equation}  \label{linearS}
{\mathcal I}(L\Lambda) = S_L {\mathcal I}(\Lambda),
\end{equation}
where $S_L$ is a $5 \times 5$ matrix with entries in ${\mathbb F}_4$.  The only nonzero entries of $S_L$ are the five entries $(S_L)_{mn}$, and they are given by  
\begin{equation}  \label{generalS}
(S_{L})_{mn} = \frac{(L \beta)_q Q_m + (L \beta)_p P_m}{\beta_q Q_n + \beta_p P_n},
\end{equation}
where, again, any choice of $\beta$ is allowed as long as the fraction is well defined.   
(We label the rows and columns of $S_L$ with integers $0, \ldots, 4$.)
A special case of Eq.~(\ref{linearS}) is the analogous equation 
for the index of a phase-space point:
\begin{equation} \label{Lonalpha}
{\mathcal I}(L\alpha) = S_L {\mathcal I}(\alpha).
\end{equation}
Let us call $S_L$ the ``index operator" for the linear transformation $L$.

As an example, consider the linear transformation we have called $H_1$:
\begin{equation}
H_1 = \mtx{cc}{1 & 0 \\ 1 & 1}.
\end{equation}
By looking at the rays of Fig.~\ref{striations}, one can see how $H_1$ maps striations to other striations.
The mapping is given by the following table.

\medskip

\begin{centering}

\begin{tabular}{c|ccccc}
$n$ & 0 & 1 & 2 & 3 & 4 \\
\hline
$m$ & 0 & 3 & 4 & 1 & 2 \\
\end{tabular}

\end{centering}

\medskip

\noindent 
Again, the only nonzero elements of $S_{H_1}$ are $(S_{H_1})_{mn}$, and we can use Eq.~(\ref{generalS}), with $L$ set equal to $H_1$, to evaluate those elements.  In this way one finds
that 
\begin{equation}
S_{H_1} = \mtx{ccccc}{1 & 0 & 0 & 0 & 0 \\ 0 & 0 & 0 & \omega & 0 \\ 0 & 0 & 0 & 0 & \bar{\omega} \\ 0 & \bar{\omega} & 0 & 0 & 0 \\ 0 & 0 & {\omega} & 0 & 0}.
\end{equation}

We now turn our attention to phase point operators.  How does the index of a phase point operator change when we apply the unitary transformation $U_L$
corresponding to a symplectic matrix $L$?  We begin with the origin operator ${\bf A}_0$ of our standard Wigner function.  For any $L$, one can simply work out what $U_L$ does to 
each of the basis vectors $|b^{(n)}_0\rangle$ of which ${\bf A}_0$ is composed.  Consider, for example, $L= H_1$, for which our chosen 
unitary transformation is (see Eq.~(\ref{UL}))
\begin{equation} 
\fixTABwidth{T}
U_{H_1} = \parenMatrixstack{0 & 0 & 0 & -i \\ 0 & 0 &1 & 0 \\ 0 & 1 & 0 & 0 \\ i & 0 & 0 & 0 } .
\end{equation}
Applying this matrix to each of the vectors $|b^{(n)}_0\rangle$ in the left-most column of Fig.~\ref{vectortable}, we see that
\begin{equation}
{\mathcal I}(U_{H_1} {\bf A}_0 U^\dag_{H_1}) = \mtx{c}{1 \\ 0 \\ 1 \\ 0 \\ \omega}.
\end{equation}
Let us label this last column vector $f_{H_1}$.  In a similar way, we can obtain an analogous vector $f_L$ for each symplectic matrix $L$.  
That is, we will have
\begin{equation} \label{aaa}
{\mathcal I}(U_{L} {\bf A}_0 U^\dag_{L}) = f_L.
\end{equation}
We call $f_L$ the ``shift vector'' for $L$.  Unlike the index operator $S_L$, the shift vector $f_L$ depends on our particular choice of the unitary $U_L$.
If instead of $U_L$ we had associated with $L$ the unitary operator $D_\beta U_L$, then instead of $f_L$ on the right-hand side of Eq.~(\ref{aaa}) we would have had the vector
$f_L + \beta_q Q + \beta_p P$.

Note that we are seeing here a difference between the action of $U_L$ on the $|b^{(n)}_0\rangle$'s and the action of $L$ on the rays $\lambda^{(n)}_0$.
Whereas $L$ merely permutes the rays among themselves, the unitary operator $U_L$---which effects the same permutation of the rows of the table---can also shift $|b^{(m)}_0\rangle$
to a different place in the row (that is, it can change the value of $k$).  This shift is precisely what we can avoid when the dimension of the Hilbert space
is an {\em odd} prime power.  But it cannot be avoided when the dimension is a power of 2.  It is the shift by $f_L$ that will force us to interpret the Wigner-function values
with respect to a different set of phase point operators.  

From the effect of $U_L$ on the index of ${\bf A}_0$, we can work out its effect on the index of any allowed phase point operator, that is,
any operator of the form (\ref{Afrombs}).  We begin with
the following relation, which is analogous to the one in Eq.~(\ref{bigone}).
\begin{equation}
\begin{split}
U_L|b^{(n)}_{\beta_q Q_n + \beta_p P_n}\rangle\langle b^{(n)}_{\beta_q Q_n + \beta_p P_n} |U^\dag_L  &= U_L D_\beta |b^{(n)}_0\rangle\langle b^{(n)}_0| D^\dag_\beta U^\dag_L \\
&= D_{L\beta} U_L |b^{(n)}_0\rangle\langle b^{(n)}_0| U^\dag_L D^\dag_{L\beta}.
\end{split}
\end{equation}
But now, in place of the last step of Eq.~(\ref{bigone}), we need to take into account the shift expressed in Eq.~(\ref{aaa}):
\begin{equation}  \label{oneterm}
U_L|b^{(n)}_{\beta_q Q_n + \beta_p P_n}\rangle\langle b^{(n)}_{\beta_q Q_n + \beta_p P_n} |U^\dag_L = D_{L\beta} |b^{(m)}_{(f_L)_m}\rangle\langle b^{(m)}_{(f_L)_m} | D^\dag_{L\beta}.
\end{equation}
As before, the $m$ in this equation labels the striation that is the image under $L$ of the striation labeled by $n$.  And $(f_L)_m$ is the $m$th component of the vector $f_L$.   
Meanwhile, Eq.~(\ref{xxx}) tells us that
\begin{equation} \label{nextequation}
D_{L\beta} |b^{(m)}_{(f_L)_m}\rangle = |b^{(m)}_{(f_L)_m + (L\beta)_q Q_m + (L\beta)_p P_m} \rangle.
\end{equation}
Now consider a phase point operator $A$ whose index is given by ${\mathcal I}(A)_n = \beta_q Q_n + \beta_p P_n$.  It follows from Eqs.~(\ref{oneterm}) and (\ref{nextequation}) that
\begin{equation}
{\mathcal I}(U_L A U^\dag_L)_m = (L\beta)_q Q_m + (L\beta)_p P_m + (f_L)_m.
\end{equation}
Recall, though, that the matrix $S_L$ is defined so as to take the index with components $\beta_q Q_n + \beta_p P_n$ to the index with components $(L\beta)_q Q_m + (L\beta)_p P_m$.  So we have
\begin{equation}  \label{LongeneralA}
{\mathcal I}(U_L A U^\dag_L) = S_L{\mathcal I}(A) + f_L.
\end{equation}
Note the difference between this formula and the analogous formula for a set of lines, Eq.~(\ref{linearS}): in Eq.~(\ref{linearS}) there is no need for the shift vector $f_L$.  

For the special case of the {\em standard} Wigner function, we can express Eq.~(\ref{LongeneralA}) in a particularly simple way.
To see how, we need to show that the vectors $|b_{k_n}^{(n)}\rangle$ used to construct ${\bf A}_\alpha$ have the same values of $k_n$ as the lines
that pass through the point $\alpha$; that is, ${\mathcal I}({\bf A}_\alpha) = {\mathcal I}(\alpha)$.   We can see this by first observing that the equation holds when $\alpha$ is the origin: the rays, which are the lines passing through the origin,
all have $k_n = 0$, and by definition, ${\bf A}_0$ is constructed from the vectors $|b_{k_n}^{(n)}\rangle$ with $k_n=0$.  We now apply a displacement $\alpha$ and use 
Eqs.~(\ref{displacelambda}) and (\ref{displaceA}):  
\begin{equation}
\begin{split}
{\mathcal I}({\bf A}_\alpha) &= {\mathcal I}({\bf A}_0) + \alpha_q Q + \alpha_p P \\
&= {\mathcal I}(\hbox{origin}) + \alpha_q Q + \alpha_p P =  {\mathcal I}(\alpha).
\end{split}
\end{equation}
So we can write
\begin{equation}
S_L{\mathcal I}({\bf A}_\alpha) = S_L{\mathcal I}(\alpha) = {\mathcal I}(L\alpha) = {\mathcal I}({\bf A}_{L\alpha}).
\end{equation}
Thus Eq.~(\ref{LongeneralA}) becomes
\begin{equation}  \label{UonstandardA}
{\mathcal I}(U_L {\bf A}_\alpha U^\dag_L) = {\mathcal I}({\bf A}_{L\alpha}) + f_L.
\end{equation}

Returning now to the case of a general phase point operator, it will be helpful to use the formula (\ref{LongeneralA}) to work out the effect of a composition of two $U_L$'s.  Suppose we 
 apply to a phase point operator $A$ the transformation $U_{L_1}$ followed by $U_{L_2}$.  The effect on the index is
 \begin{equation}  \label{composition}
 \begin{split}
  {\mathcal I}(U_{L_2}U_{L_1} A U^\dag_{L_1}U^\dag_{L_2}) &= S_{L_2} {\mathcal I}(U_{L_1} A U^\dag_{L_1}) + f_{L_2} \\
  &= S_{L_2}\left[ S_{L_1} {\mathcal I}(A) + f_{L_1} \right] + f_{L_2} \\
  &= S_{L_2}S_{L_1} {\mathcal I}(A) + S_{L_2}f_{L_1} + f_{L_2}. \\
  \end{split}
  \end{equation}
  %Thus the ``index operator'' for $U_{L_2}U_{L_1}$ is $S_{L_2}S_{L_1}$, and the corresponding 
 %R ``shift vector'' is $S_{L_2}f_{L_1} + f_{L_2}$.
 We use this formula immediately in the next subsection.

 \subsection{Finding the shift vectors $f_L$}
 
 The shift vectors are an important part of our story, so it will be good to have a simple way 
 of finding them.  First, by applying the unitaries $U_{H_x}$ of Eq.~(\ref{UL}) to the $|b^{(n)}_0\rangle$'s, we can find the four 
 shift vectors $f_{H_x}$ for $x \in {\mathbb F}_4$:
 \begin{equation} \label{fourshiftvectors0}
 H_0 = \mtx{cc}{1 & 0 \\ 0 & 1} \hspace{1cm} f_{H_0} = \mtx{c}{0 \\ 0 \\ 0 \\ 0 \\ 0} ,
 \end{equation}
 \begin{equation}  \label{fourshiftvectors1}
 H_1 = \mtx{cc}{1 & 0 \\ 1 & 1} \hspace{1cm} f_{H_1} = \mtx{c}{1  \\ 0 \\ 1 \\ 0 \\ \omega} ,
  \end{equation}
 \begin{equation}
 H_\omega = \mtx{cc}{1 & 0 \\ \omega & 1} \hspace{1cm} f_{H_\omega} = \mtx{c}{ \omega \\ 1 \\ \bar{\omega} \\ {\omega} \\ \bar{\omega}} ,
  \end{equation}
 \begin{equation} \label{fourshiftvectorswb}
 H_{\bar{\omega}} = \mtx{cc}{1 & 0 \\ \bar{\omega} & 1} \hspace{1cm} f_{H_{\bar{\omega}}} = \mtx{c}{ \bar{\omega} \\ \bar{\omega} \\ \bar{\omega} \\ \bar{\omega} \\ \bar{\omega} } .
 \end{equation}
 Recall that all the other symplectic matrices can be obtained by multiplying these four on the right and left
 by powers of the ``rotation'' $R$ (see Eq.~(\ref{RHR}))---we saw in Section II how to find $x$, $r$, and $s$ in the expression
 $L=R^r H_x R^s$.  We will use Eq.~(\ref{composition}) to see how multiplication by $R$ affects the shift vectors.  
 First note that the index operator and shift vector for the matrix $R$ are
 \begin{equation}  \label{Rindexoperator}
 S_R = \mtx{ccccc}{0 & 0 & 0 & 0 & 1 \\ 1 & 0 & 0 & 0 & 0 \\ 0 & 1 & 0 & 0 & 0 \\ 0 & 0 & 1 & 0 & 0 \\ 0 & 0 & 0 & 1 & 0 }
 \hspace{5mm} \hbox{and} \hspace{5mm}
 f_R = \mtx{c}{0 \\ 0 \\ 0 \\ 0 \\ 0}.
 \end{equation}
 Now suppose we multiply one of the $H$'s on the right by $R$.  Then according to Eq.~(\ref{composition}), with $A$ set equal to ${\bf A}_0$, we have
 \begin{equation} \label{ob1}
 f_{HR} = S_H f_R + f_H = f_H,
 \end{equation}
 since $f_R$ is the zero vector.  So multiplying on the right by $R$ does not change the shift vector.  
 Now suppose we multiply one of the $H$'s on the left by $R$.  We have
 \begin{equation} \label{ob2}
 f_{RH} = S_R f_H + f_R = S_R f_H.
 \end{equation}
 That is, we cyclically shift $f_H$ one step downward to get $f_{RH}$.  
 %CHANGE HERE
 %(Eqs.~(\ref{ob1}) and (\ref{ob2}) can also be derived from the definition of the $|b^{(n)}_k\rangle$'s.)
 Continuing in this way, we can find the shift vector for each of our 60 symplectic matrices as expressed
 in Eq.~(\ref{RHR}). 
 These 60 matrices can be sorted into 12 sets of 5 each, such that the matrices within each set all have the same
 shift vector (since multiplying on the right by $R$ does not change $f$).  Thus for our 60 $L$'s, there are only 12 distinct
 shift vectors $f_L$.  
 
 In fact we can obtain a formula that yields $f_L$ directly for any $L$, bypassing the explicit decomposition
 of $L$ into the form $R^r H_x R^s$.  We begin by noting that the following 12 vectors constitute the full
 set of nonzero vectors $\mu$ in ${\mathbb F}_4^2$ whose slopes $\mu_p/\mu_q$ are not infinite.
 \begin{equation} \label{notinfinite}
 \begin{split}
 &H_x H_{\bar{\omega}}^T H_x^T \mtx{c}{1 \\ 0}, \;\; x=0\; \hbox{or} \;\bar{\omega} ,\\
 &R^t H_x H_{\bar{\omega}}^T H_x^T R^t \mtx{c}{1 \\ 0}, \;\; x=1\; \hbox{or} \;{\omega} ,
 \end{split}
 \end{equation}
 where $t = 0, \ldots, 4$.
 Now, for any symplectic matrix $L$ and for $n=0, \ldots, 4$, consider the vector $\mu^{(n)}_L$ defined by
 \begin{equation}  \label{f1}
 \mu^{(n)}_L = R^{-n} L H_{\bar{\omega}}^T L^T R^{-n} \mtx{c}{1 \\ 0}.
 \end{equation}
 The slope of this vector---which is not infinite since $\mu^{(n)}_L$ reduces to one of the vectors in Eq.~(\ref{notinfinite})---turns out to be equal to the value of $(f_L)_n$:
 \begin{equation}  \label{f2}
 (f_L)_n =  \frac{\big(\mu_L^{(n)}\big)_p}{ \big(\mu^{(n)}_L\big)_q}  .
 \end{equation}
 To see this, one can first check that the relation agrees with Eqs.~(\ref{fourshiftvectors0})--(\ref{fourshiftvectorswb}) when $L = H_x$ with $x \in {\mathbb F}_4$.
 From there it suffices to make sure the formula behaves correctly when we multiply $L$ on the left or right
 by $R$.  But this is indeed the case:
 \begin{equation}
 \begin{split}
  &\mu^{(n)}_{LR} =  \mu^{(n)}_L \hspace{1mm} \hbox{since $RH_{\bar{\omega}}^TR = H_{\bar{\omega}}^T$ }, \\
  &\mu^{(n)}_{RL} = R^{-n+1} L H_{\bar{\omega}}^T L^T R^{-n+1} \mtx{c}{1 \\ 0} = \mu^{(n-1)}_{L},
  \end{split}
  \end{equation}
  where the superscript on $\mu$ is to be interpreted mod 5.
  Thus Eqs.~(\ref{f1}) and (\ref{f2}) constitute a valid formula for $f_L$.

\section{Reinterpreting Wigner-function values}

We now ask how the effect of $U_L$ is to be described in phase space.  
Let the initial state of our two qubits be $\rho$, and let us express this state as a Wigner function, adopting for now
the ``standard'' definition ${\bf W}$:
 \begin{equation}
 {\bf W}_\alpha(\rho) = \tfrac{1}{4} \hbox{Tr}({\bf A}_\alpha \rho).
 \end{equation}
We perform on $\rho$ one of our unitary transformations $U_L$.  
What does the action of $U_L$ look like in phase space?

Let $\rho'$ be the transformed density matrix: $\rho' = U_L \rho U^\dag_L$.  Then $\rho=U^\dag_L \rho' U_L$, 
and we can write
\begin{equation}  \label{abc}
{\bf W}_\alpha(\rho) = \tfrac{1}{4} \hbox{Tr}({\bf A}_\alpha U^\dag_L \rho' U_L) 
= \tfrac{1}{4} \hbox{Tr}\left[ (U_L{\bf A}_\alpha U^\dag_L) \rho' \right].
\end{equation}
%Now, the index of $U_L{\bf A}_\alpha U^\dag_L$ is 
%\begin{equation}
%{\mathcal I}(U_L{\bf A}_\alpha U^\dag_L) = {\mathcal I}({\bf A}_{L\alpha}) + f_L.
%\end{equation}
Let us define the ``$f$-Wigner function'' $W^{f}_\alpha$ such that
it has as its origin operator not ${\bf A}_0$, whose index is the zero vector, but rather the operator whose index is $f$:
\begin{equation}
A^{f}_0 = \sum_n |b^{(n)}_{f_n}\rangle \langle b^{(n)}_{f_n} | - I.
\end{equation}
The other phase point operators are defined in the usual way:
\begin{equation}
A^{f}_\alpha = D_\alpha A^{f}_0 D^\dag_\alpha,
\end{equation}
and the $f$-Wigner function itself is given by
\begin{equation}
W^{f}_\alpha = \tfrac{1}{4}\hbox{Tr}(A^{f}_\alpha \rho).
\end{equation}
Let us now examine the index of the operator in parentheses on the right-hand side of Eq.~(\ref{abc}).
\begin{equation}  \label{specialproof}
\begin{split}
{\mathcal I}(U_L{\bf A}_\alpha U^\dag_L) &=S_L {\mathcal I}({\bf A}_\alpha) + f_L \\
&= S_L(\alpha_q Q + \alpha_p P) + f_L \\
&=(L\alpha)_q Q + (L\alpha)_p P + f_L \\
&={\mathcal I}(A^{f_L}_{L\alpha}).
\end{split}
\end{equation}
Thus the operator in parentheses on the right-hand side of Eq.~(\ref{abc}) is $A^{f_L}_{L\alpha}$, and the
trace in that equation gives us the Wigner function $W^{f_L}_{L\alpha}(\rho')$.  So we have
\begin{equation}
{\bf W}_\alpha(\rho) = W^{f_L}_{L\alpha}(U_L \rho U^\dag_L).
\end{equation}
We can write this equation in a form analogous to that of Eq.~(\ref{Wignermove}) or Eq.~(\ref{reinterpret}):
\begin{equation}  \label{mainresult}
W^{f_L}_\alpha(U_L \rho U^\dag_L) = {\bf W}_{L^{-1}\alpha}(\rho).
\end{equation}
That is, performing the unitary $U_L$ is equivalent to moving the values of the Wigner function
in phase space according to $L$, {\em and} reinterpreting those values as values of the 
new Wigner function $W^{f_L}$.  

As an example, let us consider the state $\rho=|\psi\rangle\langle\psi|$, where $|\psi\rangle$ is the product vector
\begin{equation}  \label{examplerho}
|\psi\rangle = \mid\uparrow\rangle\otimes\hspace{-0.8mm}\mid\rightarrow\rangle = \mtx{c}{1 \\ 0} \otimes \frac{1}{\sqrt{2}} \mtx{c}{1 \\ 1} = \frac{1}{\sqrt{2}}\mtx{c}{1 \\ 1 \\ 0 \\ 0}.
\end{equation}
Here I am using arrows to represent standard Bloch-sphere vectors.  In our standard Wigner-function representation, 
this state appears as

\bigskip

\begin{centering}

${\bf W}(\rho) = \frac{1}{4}$
\setlength{\tabcolsep}{2pt}
 \hspace{-2mm}
\begin{tabular}{|c|c|c|c|c}
\hline
$\,\, 1\,\,$ & $\,\, 0\,\,$ & $\,\, 1 \,\,$ & $\,\, 0 \,\,$  \\
\hline
0 & 0 & 0 & 0  \\
\hline
0 & 0 & $0$ & $0$  \\
\hline
$\,\, 1\,\,$ & $\,\, 0\,\,$ & $1$ & $0$  \\
\hline
\end{tabular}
\begin{tabular}{c}
$\leftarrow\rightarrow$ \\
$\rightarrow\leftarrow$ \\
$\leftarrow\leftarrow$ \\
$\rightarrow\rightarrow$
\end{tabular}

\hspace{5.4mm}
\setlength{\tabcolsep}{3.1pt}
\begin{tabular}{cccc}
$\uparrow\uparrow$ & $\downarrow\downarrow$ & $\uparrow\downarrow$ & $\downarrow\uparrow$ 
\end{tabular}

\bigskip

\end{centering}

\noindent The arrows give the interpretation of the vertical and horizontal lines, in accordance with the values 
$k_0 = 0$ and $k_1 = 0$ for the standard Wigner function.  Note that the sums over the vertical and horizontal lines
give the correct probabilities of the states associated with those lines.  (The same holds for the other lines of phase space, which are not labeled in the diagram.)

We now perform on $\rho$ the unitary transformation $U_G$ corresponding to the symplectic matrix
\begin{equation}
G = R^2 H_1 R = \mtx{cc}{\bar{\omega} & 0 \\ 0 & \omega}.  
\end{equation}
(See Eqs.~(\ref{exampleL}) and (\ref{exampleUL}).)
The resulting state is $\rho' = |\psi'\rangle\langle\psi'|$, where $|\psi'\rangle$ is the maximally entangled state
\begin{equation}  \label{examplestate1}
|\psi'\rangle = \frac{1}{\sqrt{2}}(\mid\downarrow\rangle \otimes\hspace{-0.8mm} \mid\downarrow\rangle - \mid\uparrow\rangle \otimes\hspace{-0.8mm} \mid\uparrow\rangle) = \frac{1}{\sqrt{2}}\mtx{c}{-1 \\ 0 \\ 0 \\ 1}.
\end{equation}
We would like to express this state as a Wigner function, using the definition $W^{f_G}_\alpha$.  For the matrix $G$, we can find $f_G$ either by starting with 
$f_{H_1}$, which we found earlier, and shifting it downward by two steps (since $G$ has two factors of $R$ on the left of $H_1$) or by using
the formula in Eqs.~(\ref{f1}) and (\ref{f2}).  The result is
\begin{equation}
f_G = \mtx{c}{0 \\ \omega \\ 1 \\ 0 \\ 1}.
\end{equation}
This $f_G$ tells us how to define the origin operator $A^{f_G}_0$, and the other phase point operators are obtained via displacements.  
One finds that the $f_G$-Wigner function of $\rho'$ is

\bigskip

\begin{centering}

${W}^{f_G}(\rho') = \frac{1}{4}$
\setlength{\tabcolsep}{2pt}
 \hspace{-2mm}
\begin{tabular}{|c|c|c|c|c}
\hline
$\,\, 0\,\,$ & $\,\, 0\,\,$ & $\,\, 0 \,\,$ & $\,\, 0 \,\,$  \\
\hline
0 & 0 & 0 & 0  \\
\hline
1 & 1 & $0$ & $0$  \\
\hline
$\,\, 1\,\,$ & $\,\, 1\,\,$ & $0$ & $0$  \\
\hline
\end{tabular}
\begin{tabular}{c}
$\leftarrow\leftarrow$ \\
$\rightarrow\rightarrow$ \\
$\leftarrow\rightarrow$ \\
$\rightarrow\leftarrow$ 
\end{tabular}

\hspace{9.5mm}
\setlength{\tabcolsep}{3.1pt}
\begin{tabular}{cccc}
$\uparrow\uparrow$ & $\downarrow\downarrow$ & $\uparrow\downarrow$ & $\downarrow\uparrow$  
\end{tabular}

\bigskip

\end{centering}

\noindent Note that the horizontal lines now have a different interpretation than before; this is because the $n=1$ component of $f_G$ 
is not zero.  In effect, the labels of the horizontal lines have been displaced by $\omega$, since $(f_G)_1 = \omega$.  One can check that the sum of the Wigner function over each vertical and horizontal line yields the correct probability of the outcome associated with that line
when the system is in the state $\rho'$.  

One can also check that the locations of the nonzero entries in the above table are indeed the images of the locations of the nonzero entries in the earlier 
table under the operation of the symplectic matrix $G$.  In this way, the example illustrates Eq.~(\ref{mainresult}).   (Again, we never
change the order of the field elements $0,1,\omega,\bar{\omega}$ labeling the two axes.  It is only the physical interpretation of those
values that changes.)

So far in this section, we have expressed our initial state as a {\em standard} Wigner function ${\bf W}_\alpha$.  
But suppose we are starting by expressing our initial state $\rho$ in the \hbox{$f$-Wigner} framework for some index $f$.  
If we now perform $U_L$ on $\rho$ to get $\rho' = U_L \rho U^\dag_L$, what does this transformation look like
in phase space?  

We now have, in place of Eq.~(\ref{abc}), the equation
\begin{equation}  \label{newstart}
W^f_\alpha(\rho) = \tfrac{1}{4}\hbox{Tr}\left[ (U_L A^f_\alpha U^\dag_L) \rho' \right].
\end{equation}
As before, we wish to re-express the operator in parentheses as an operator
belonging to a different definition of the Wigner function.  We can generalize
Eq.~(\ref{specialproof}) as follows.
\begin{equation}  \label{generalproof}
\begin{split}
{\mathcal I}(U_L A_\alpha^f U^\dag_L) &= S_L{\mathcal I}(A_\alpha^f) + f_L \\
&=S_L(f + \alpha_q Q + \alpha_p P) + f_L \\
&=S_L f  + (L\alpha)_q Q + (L\alpha)_p P + f_L \\
&={\mathcal I}(A^{S_L f + f_L}_{L\alpha}).
\end{split}
\end{equation}
So the operator in parentheses on the right-hand side of Eq.~(\ref{newstart}) is
$A^{S_L f + f_L}_{L\alpha}$, and that equation can thus be rewritten as
\begin{equation}
W^f_\alpha(\rho) = W^{S_L f + f_L}_{L\alpha}(U_L \rho U^\dag_L).
\end{equation}
The form of the equation analogous to Eqs.~(\ref{Wignermove}), (\ref{reinterpret}), and (\ref{mainresult}) is
\begin{equation} \label{realmainresult}
W^{S_L f + f_L}_\alpha(U_L \rho U^\dag_L) = W^f_{L^{-1}\alpha}(\rho).
\end{equation}
So if we are already in the $f$ representation and perform the unitary $U_{L}$, we let $L$ act on the 
points of phase space {\em and} we change our representation from $f$ to $S_{L}f + f_{L}$.

To see how this works, let us go back to the above example.  Again, we started with the state
$|\psi\rangle = \mid\uparrow\rangle \otimes\hspace{-0.8mm} \mid \rightarrow\rangle$ and applied the unitary operator $U_G$ of Eq.~(\ref{exampleUL}).
The result was an entangled state.  We now take this entangled state as our starting state and rename it $\rho$.  We copy its $W^{f_G}$ representation here:

\bigskip

\begin{centering}

${W}^{f_G}(\rho) = \frac{1}{4}$
\setlength{\tabcolsep}{2pt}
 \hspace{-2mm}
\begin{tabular}{|c|c|c|c|c}
\hline
$\,\, 0\,\,$ & $\,\, 0\,\,$ & $\,\, 0 \,\,$ & $\,\, 0 \,\,$  \\
\hline
0 & 0 & 0 & 0  \\
\hline
1 & 1 & $0$ & $0$  \\
\hline
$\,\, 1\,\,$ & $\,\, 1\,\,$ & $0$ & $0$  \\
\hline
\end{tabular}
\begin{tabular}{c}
$\leftarrow\leftarrow$ \\
$\rightarrow\rightarrow$ \\
$\leftarrow\rightarrow$ \\
$\rightarrow\leftarrow$ 
\end{tabular}

\hspace{8.9mm}
\setlength{\tabcolsep}{3.1pt}
\begin{tabular}{cccc}
$\uparrow\uparrow$ & $\downarrow\downarrow$ & $\uparrow\downarrow$ & $\downarrow\uparrow$  
\end{tabular}

\bigskip

\end{centering}

\noindent Suppose we now apply $U_G$ a second time.  The resulting state is 
\begin{equation} \label{examplestate2}
U_G \left[ \frac{1}{\sqrt{2}}(\mid\downarrow\rangle \otimes\hspace{-0.8mm} \mid\downarrow\rangle - \mid\uparrow\rangle \otimes\hspace{-0.8mm} \mid\uparrow\rangle)\right] = \;  \mid\leftarrow\rangle \otimes \hspace{-0.6mm} \mid\uparrow\rangle = \frac{1}{\sqrt{2}}\mtx{c}{1 \\ 0 \\ -1 \\ 0}.
\end{equation}
Let $\rho'$ be the density matrix representing this state.  
According to Eq.~(\ref{realmainresult}), we should now use the Wigner representation associated with the index
$S_{L}f + f_{L}$.  In our present example, this index is $S_Gf_G + f_G$, which works out to be 
\begin{equation}    \label{weirdindex}
S_Gf_G + f_G = \mtx{c}{0 \\ 1 \\ 0  \\ \omega \\ 1}.
\end{equation}
(Alternatively, we can observe that $G^2=R^4 H_1 R^3$ and find $f_{G^2}$ starting with Eq.~(\ref{fourshiftvectors1}).)
With this definition of the Wigner function, one finds that

\bigskip

\begin{centering}

${W}^{S_Gf_G+f_G}(\rho') = \frac{1}{4}$
\setlength{\tabcolsep}{2pt}
 \hspace{-2mm}
\begin{tabular}{|c|c|c|c|c}
\hline
$\,\, 0\,\,$ & $\,\, 0\,\,$ & $\,\, 0 \,\,$ & $\,\, 0 \,\,$  \\
\hline
1 & 0 & 0 & 1  \\
\hline
0 & 0 & $0$ & $0$  \\
\hline
$\,\, 1\,\,$ & $\,\, 0\,\,$ & $0$ & $1$  \\
\hline
\end{tabular}
\begin{tabular}{c}
$\rightarrow\leftarrow$ \\
$\leftarrow\rightarrow$ \\
$\rightarrow\rightarrow$ \\
$\leftarrow\leftarrow$
\end{tabular}

\hspace{18.9mm}
\setlength{\tabcolsep}{3.1pt}
\begin{tabular}{cccc}
$\uparrow\uparrow$ & $\downarrow\downarrow$ & $\uparrow\downarrow$ & $\downarrow\uparrow$ 
\end{tabular}

\bigskip

\end{centering}

\noindent Note that the arrow labels of the horizontal lines have again changed, because the $n=1$ component
of the index has changed.  This is part of the reinterpretation of the Wigner-function values.  The labels are
consistent, though, with the state.  For example, the probability of finding the first qubit in the state $\mid\leftarrow\rangle$,
as computed by summing the Wigner-function over the rows labeled $\leftarrow\leftarrow$ and $\leftarrow\rightarrow$, is one, as it should be.
This example illustrates Eq.~(\ref{realmainresult}): the locations of the nonzero entries in the table for $\rho'$
are the images, under the symplectic transformation $G$, of the locations of the nonzero entries in the table for $\rho$. 

Suppose now that, beginning with a state expressed in the standard Wigner function, we execute a series of
unitary operations $U_{L_1}, U_{L_2}, \ldots, U_{L_N}$.  We will typically run through a series of $f$-Wigner functions,
where the various $f$-values arise from repeated applications of the formula $f \rightarrow S_L f + f_L$. 
Or, if we have kept track of the $L_j$'s, then at the $M$th stage, we use the Wigner function defined by
$f_{L_M \cdots L_1}$.  For any such string of unitaries, the only Wigner-function definitions we will ever need are the
12 $f$-Wigner functions associated with our 12 possible shift vectors $f_L$.

\section{Discussion}

We began this paper by recalling a remarkable fact about discrete Wigner functions in odd prime-power dimensions: for every symplectic linear transformation $L$
on the discrete phase space, there exists a unitary operator $U_L$ corresponding to $L$ such that
\begin{equation}
W_\alpha(U_L \rho U_L^\dag) = W_{L^{-1}\alpha}(\rho).
\end{equation}
That is, the effect of this unitary is simply to permute the values of the Wigner function in accordance with the linear transformation $L$.
Our work in this paper was motivated by the fact that there is no such equation for a $d \times d$ phase space when the dimension $d$ is a power of two.  

We have found that there {\em is} a similar equation for the case $d=4$, but it requires reinterpreting the values of the Wigner function with respect to
a variety of operator bases.  We can write the equation as
\begin{equation}  \label{newcovariance}
W^g_\alpha(U_L \rho U_L^\dag) = W^f_{L^{-1}\alpha}(\rho),
\end{equation}
where $g$ and $f$ are five-component vectors with entries in ${\mathbb F}_4$.  
The index $g$ depends on both $f$ and $L$ in accordance with Eq.~(\ref{realmainresult}): $g = S_L f + f_L$.
We have noted that the full set of indices one will need consists of twelve shift vectors, given by Eqs.~(\ref{fourshiftvectors0})-(\ref{fourshiftvectorswb}) together
with the shift rule associated with multiplication on the left by $R$.    

The number twelve as it appears here ties into one of the results of Ref.~\cite{Gibbons}, as we now explain.

Ref.~\cite{Gibbons} classifies the allowed definitions of the discrete Wigner function for a given dimension---or one could say it classifies the allowed
choices of the set of phase point operators $\{A_\alpha\}$---into ``equivalence classes'' and ``similarity classes.''  Two sets of operators $\{A_\alpha\}$ and $\{A_\alpha'\}$ 
are {\em equivalent} if they are related to each other by displacement: $A_\alpha' = D_\beta A_\alpha D_\beta^\dag$.  They are {\em similar}
if they are related to each other by a symplectic linear transformation and an associated unitary: $A_{L\alpha}' = UA_\alpha U^\dag$.  
Comparing this last relation with Eq.~(\ref{generalproof}), we see that all the sets of phase point operators $\{A^f_\alpha\}$ that we will ever need in the reinterpretation 
scheme of the preceding section belong to the same similarity class.  

One can also find in Ref.~\cite{Gibbons} a formula, for the case $d=4$, that tells us to which similarity class any legitimate set of phase point operators belongs.  This formula
is a function whose input is essentially the {\em index} of $A_0$ (though the conventions in that paper are different) and whose output is 
an element of ${\mathbb F}_4$ that labels the similarity class.  (There are exactly four similarity classes when $d=4$.)  The similarity-class value in that paper 
was called $D$ (not to be confused with displacement operators), and it happens that all of the $f$-Wigner functions emerging from the construction
described in the preceding section
belong to the class $D=\bar{\omega}$.  It was noted in Ref.~\cite{Gibbons} that this particular similarity class consists of exactly twelve equivalence classes.
These twelve correspond to our twelve canonical shift vectors $f_L$.  Each of the associated $f$-Wigner functions is in a different equivalence class.  

It is worth rewriting the formula for the similarity-class value $D$ in terms of the index convention we have introduced in the present paper.  
And in order to label the similarity classes in a way that privileges the
particular class we have been using here, let us define $E$ to be $D + \bar{\omega}$.  Then the similarity class considered in this paper
is the one with $E=0$.  Writing down the formula for $E$ is a matter of starting with the formula for $D$ in Ref.~\cite{Gibbons} and re-expressing it
in terms of the $k_n$'s of Fig.~\ref{vectortable}.  One finds that for any shift vector $f$---not necessarily one that arises in our reinterpretation 
scheme---the similarity class of the $f$-Wigner function has the $E$-value given by
\begin{equation} \label{Cform}
E(f) = r^T {f} + {f}^T M {f},
\end{equation}
where
\begin{equation}
r = {\omega}\mtx{c}{1 \\ 1 \\ 1 \\ 1 \\ 1} \hspace{6mm}\hbox{and}\hspace{6mm}
M = \mtx{ccccc}{0 & 1 & {\omega} & 0 & 0 \\ 0 & 0 & 1 & {\omega} & 0 \\ 0 & 0 & 0 & 1 & {\omega} \\ {\omega} & 0 & 0 & 0 & 1 \\ 1 & {\omega} & 0 & 0 & 0}.
\end{equation}
%The vector $r$ in this formula is unique, but there are many ways we could have written the matrix $M$: any two elements $M_{jk}$ and $M_{kj}$ (one of which is zero)
%could be interchanged without affecting the value of $C$.  
Clearly our standard Wigner function, whose $f$ vector consists entirely of zeroes, has $E = 0$.   It must also be the case that for any symplectic matrix $L$, the
transformation $f \rightarrow S_L f + f_L$ preserves the value of $E$.  (Again we are comparing Eq.~(\ref{generalproof}) to the definition of ``similar.'')  In particular, for any $L$, we have $E(f_L) = 0$.

Finally, we note that all the $f$-Wigner functions in the similarity class $E=0$ have a special kind of symmetry.  
Consider first our standard Wigner function ${\bf W}$.  From Eq.~(\ref{mainresult}) with $L$ set equal to $R$, we find that
our standard Wigner function is covariant with respect to the unitary ``rotation'' operator $U_R$:
\begin{equation}  \label{ppp1}
{\bf W}_\alpha(U_R \rho U^\dag_R) = {\bf W}_{R^{-1}\alpha}(\rho),
\end{equation}
where we have used the fact that $f_R$ is the zero vector.
To see how this relation generalizes to the $f_L$-Wigner function, we begin again with Eq.~(\ref{mainresult}):
\begin{equation}  \label{ppp2}
W^{f_L}_\alpha(U_L \rho U^\dag_L) = {\bf W}_{L^{-1}\alpha}(\rho).
\end{equation}
From Eqs.~(\ref{ppp1}) and (\ref{ppp2}), it follows that
\begin{equation}
W_\alpha^{f_L}(V_{R_L}\rho V^\dag_{R_L}) = W^{f_L}_{R_L^{-1}\alpha}(\rho),
\end{equation}
where 
\begin{equation}
R_L = LRL^{-1} \hspace{3mm}\hbox{and}\hspace{3mm} V_{R_L} = U_L U_R U^\dag_L.
\end{equation}
The matrix $R_L$ and the unitary $V_{R_L}$ have the following two important properties:
(i) $V_{R_L}$ is a unitary operator that we can associate with $R_L$, in the sense that
\begin{equation}
V_{R_L} D_\beta V^\dag_{R_L} \propto D_{R_L \beta},
\end{equation}
and (ii) the matrix $R_L$ has as much right to be regarded as a rotation matrix as $R$ does, in that it 
has period five and cycles through all five striations.  
Thus, the $f_L$-Wigner function has a kind of rotational symmetry, just as ${\bf W}$ does, but with a different
notion of rotation.  Moreover, one can extend this argument to any other $f$-Wigner function with $E(f)=0$, using the fact 
that each such $f$ can be obtained from one of the $f_L$'s by a translation.  

Indeed, we could have built analogues of Figures 1 and 2 based on the special rotation matrix and special ``rotation unitary'' associated with any of these $f$-Wigner functions,
in which case that particular $f$-Wigner function would have emerged as the natural ``standard'' Wigner function.   
So it would be wrong to suppose that our work in this paper favors a single definition of the Wigner function
for a pair of qubits. What it does, rather, is to show how one can select a relatively small set of such definitions---exactly twelve---that collectively exhibit a kind of covariance under the restricted Clifford group.

\section*{Acknowledgements}
I thank Billy Braasch for a stimulating conversation that sparked the research presented in this paper.

%\begin{equation}  \label{rcondition4}
%\end{equation}

\section*{Appendix}

Here we show that the unitary matrices $U_L$ defined in Section II satisfy Eq.~(\ref{rep}).  That is, we show that
\begin{equation} \label{app1}
U_{L_1}U_{L_2} \propto U_{L_1L_2}
\end{equation}
for all pairs of symplectic matrices $L_1$ and $L_2$.

We begin with a few special cases that are easily checked:
\begin{equation} \label{facts}
\begin{split}
& U_{H_x}U_{H_y} = U_{H_x H_y}, \; \hbox{for} \; x,y \in {\mathbb F}_4, \\
& U_R U_{H_{\bar{\omega}}} U_R = U_{R H_{\bar{\omega}} R} = U_{H_{\bar{\omega}}} , \\ 
& U_R^5 = U_{R^5} = U_{H_0} = I.
\end{split}
\end{equation}

We now claim that it is sufficient to show that 
\begin{equation}  \label{app2}
U_{H_x} U_R^s U_{H_y} \propto U_{H_x R^s H_y}, \; \hbox{for} \; x,y \in {\mathbb F}_4, \; s=0, \ldots, 4.
\end{equation}
It will then immediately follow that Eq.~(\ref{app1}) is true for $L_1$ and $L_2$ of the forms
\begin{equation}
L_1 = H_{x} R^{s_1} \hspace{5mm} \hbox{and} \hspace{5mm} L_2 = R^{r_2}H_{y}.
\end{equation}
Extending this result to all $L_1$ and $L_2$ is a matter of multiplying the above $L_1$ on the left by $R^{r_1}$, multiplying $L_2$ on the
right by $R^{s_2}$, and using the fact that $U_{R^r H_x R^s}$ is defined to be $U_R^r U_{H_x} U_R^s$.  

So we turn our attention to demonstrating Eq.~(\ref{app2}).

If either $x$ or $y$ is zero, the equation reduces to the definition of $U_{R^r H_x R^s}$ just quoted.  

Now suppose that $x=\bar{\omega}$.
Then on the left-hand side of Eq.~(\ref{app2}), we can write
\begin{equation}
U_{H_{\bar{\omega}}} U_R^s = U_R^{-s}\big(U_R^s U_{H_{\bar{\omega}}} U_R^s \big) = U_R^{-s} U_{H_{\bar{\omega}}},
\end{equation}
so that $U_{H_{\bar{\omega}}}$ directly multiplies $U_{H_y}$ and we can combine the two unitaries as in Eq.~(\ref{facts}).
Meanwhile, on the right-hand side of Eq.~(\ref{app2}), we can similarly remove the $R$'s lying between $H_{\bar{\omega}}$ and $H_y$ and
combine those two matrices into a single $H$ matrix.  And then Eq.~(\ref{app2}) again follows from the definition of $U_{R^r H_x R^s}$. An analogous
argument applies when $x$ has an arbitrary value and $y$ is equal to $\bar{\omega}$.  

This leaves us with the cases where $(x,y)$ is equal to $(1,1)$, $(1, \omega)$, $(\omega, 1)$, and $(\omega, \omega)$.  In fact, though, it is enough to 
deal with the case $(x,y) = (1,1)$, for which Eq.~(\ref{app2}) becomes
\begin{equation}  \label{app3}
U_{H_1} U_R^s U_{H_1} \propto U_{H_1 R^s H_1}.
\end{equation}
By multiplying both sides of this equation on the right by $U_{H_{\bar{\omega}}}$ and using essentially the same argument as in the preceding paragraph, 
we can turn this equation into
\begin{equation}
U_{H_1} U_R^s U_{H_\omega} \propto U_{H_1 R^s H_\omega}.
\end{equation}
And we can make the same transformation on the left-hand sides of the expressions.  All that remains, then, is to check that Eq.~(\ref{app3}) holds for 
$s=0, \ldots, 4$.  Upon checking directly, one finds that the equation indeed holds.


\begin{thebibliography}{99}

\bibitem{Hillery} M.~Hillery, R.~F.~O’Connell, M.~O.~Scully, and E.~P.~Wigner, Distribution functions in physics: Fundamentals, {\em Phys. Rep.} {\bf 106} (1984) 123.
\bibitem{Weinbub} J.~Weinbub and D.~K.~Ferry, Recent advances in Wigner distribution approaches, {\em App.~Phys.~Rev.} {\bf 5} (2018) 041104.  
%\bibitem{Littlejohn} R.~G.~Littlejohn, The semiclassical evolution of wave packets, {\em Phys.~Rep.} {\bf 138} (1986) 193.
\bibitem{Arvind} Arvind, B.~Dutta, N.~Mukunda, and R.~Simon, The Real Symplectic Groups in Quantum Mechanics and Optics, {\em Pramana} {\bf 45} (1995) 471.
\bibitem{Ferraro} A.~Ferraro, S.~Olivares, and M.~G.~A.~Paris, Gaussian states in continuous variable quantum information, arxiv:quant-ph/0503237 (2005).
\bibitem{Gibbons} K.~S.~Gibbons, M.~J.~Hoffman, and W.~K.~Wootters, Discrete phase space based on finite fields, {\em Phys.~Rev.~A} {\bf 70} (2004) 062101.
\bibitem{Klimov} A.~B.~Klimov and C.~Mu{\~n}oz, Discrete Wigner function dynamics, {\em J.~Opt.~Soc.~B} {\bf 7} (2005) S588.
\bibitem{Vourdas} A.~Vourdas, Galois quantum systems, {\em J.~Phys.~A: Math.~Gen.} {\bf 38} (2005) 8453.
\bibitem{Gross} D.~Gross, Hudson's theorem for finite-dimensional quantum systems, {\em J.~Math.~Phys.} {\bf 47} (2006) 122107.
\bibitem{Bengtsson} I.~Bengtsson and K.~\.{Z}yczkowski, On discrete structures in finite Hilbert space, arxiv:1701.07902 (2017).
\bibitem{Zhu3} H.~Zhu, Permutation Symmetry Determines the Discrete Wigner Function, {\em Phys.~Rev.~Lett.} {\bf 116} (2016) 040501.
\bibitem{Feynman} R.~P.~Feynman, Negative probability, in {\em Quantum Implications: Essays in Honour of David Bohm}, edited by B.~Hiley and D.~Peat (Routledge, London, 1987).
\bibitem{Wootters} W.~K.~Wootters, A Wigner-function formulation of finite-state quantum mechanics, {\em Ann.~Phys.~(N.Y.)} {\bf 176} (1987) 1.
\bibitem{Zhu} H.~Zhu, Multiqubit Clifford groups are unitary 3-designs, {\em Phys.~Rev.~A} {\bf 96} (2017) 062336.
\bibitem{Raus1} R.~Raussendorf, D.~E.~Browne, N.~Delfosse, C.~Okay, and J.~Bermejo-Vega, Contextuality and Wigner function negativity in qubit quantum computation, {\em Phys.~Rev.~A} {\bf 95} (2017) 0523334.
\bibitem{Feldmann} R.~Raussendorf, C.~Okay, M.~Zurel, and P.~Feldmann, The role of cohomology in quantum computation with magic states, {\em Quantum} {\bf 7} (2023) 979.
\bibitem{Raussendorf2} R.~Raussendorf, J.~Bermejo-Vega, E.~Tyhurst, C.~Okay, and M.~Zurel, Phase-space-simulation method for quantum computation with magic states on qubits, {\em Phys.~Rev.~A} {\bf 101} (2020) 012350.
\bibitem{Kocia} L.~Kocia and P.~Love, Discrete Wigner Formulation for Qubits and Non-Contextuality of Clifford Gates on Qubit Stabilizer States, {\em Phys.~Rev.~A} {\bf 96} (2017) 062134.
\bibitem{Leonhardt1} U.~Leonhardt, Quantum-State Tomography and Discrete Wigner Function, {\em Phys.~Rev.~Lett.} {\bf 74} (1995) 4101.
\bibitem{Leonhardt2} U.~Leonhardt, Discrete Wigner function and quantum-state tomography, {\em Phys.~Rev.~A} {\bf 53} (1996) 2998.
\bibitem{Hashimoto} T.~Hashimoto, M.~Horibe, and A.~Hayashi, Unitary Representation of Symplectic Group for Phase Point Operators in Discrete Phase Space, {\em J.~Phys.~A: Math.~Th.} {\bf 40} (2007) 14253.
\bibitem{Veitch1} V.~Veitch, C.~Ferrie, D.~Gross, and J.~Emerson, Negative quasi-probability as a resource for quantum computation, {\em New J.~Phys.} {\bf 14} (2012) 113011.
\bibitem{Howard} M.~Howard, J.~Wallman, V.~Veitch, and J.~Emerson, Contextuality supplies the `magic' for quantum computation, {\em Nature} {\bf 510} (2014) 351.
\bibitem{Veitch2} V.~Veitch, S.~H.~Mousavian, D.~Gottesman, and J.~Emerson, The resource theory of stabilizer quantum computation, {\em New J.~Phys.} {\bf 16} (2014) 013009.
\bibitem{Delfosse} N.~Delfosse, P.~A.~Guerin, J.~B, and R.~Raussendorf, Wigner function negativity and contextuality in quantum computation on rebits, {\em Phys.~Rev.~X} {\bf 5} (2015) 021003.
%\bibitem{Raussendorf} R.~Raussendorf, D.~E.~Browne, N.~Delfosse, C.~Okay, and J.~Bermejo-Vega, Contextuality and Wigner function negativity in qubit quantum computation, {\em Phys.~Rev.~A} {\bf 95} (2017) 052334.
\bibitem{Liu} Z.-W.~Liu and A.~Winter, Many-body quantum magic, {\em PRX Quantum} {\bf 3} (2022) 020333.
\bibitem{Galetti} D.~Galetti and A.~F.~R.~De Toledo Piza, An extended Weyl-Wigner transformation for special finite spaces, {\em Physica A} {\bf 149} (1988) 267.


\bibitem{Zhu1} H.~Zhu, Mutually unbiased bases as minimal Clifford covariant 2-designs, {\em Phys.~Rev.~A} {\bf 91} (2015) 060301.

\bibitem{Appleby} D.~M.~Appleby, Properties of the extended Clifford group with applications to SIC-POVMs and MUBs, arxiv:0909.5233 (2004).
\bibitem{Galvao} E.~F.~Galv{\~a}o, Discrete Wigner functions and quantum computational speedup, {\em Phys.~Rev.~A} {\bf 71} (2005) 042302.
\bibitem{Paz1} C~Cormick, E.~F.~Galv{\~a}o, D.~Gottesman, J.~P.~Paz, and A.~O.~Pittenger, Classicality in discrete Wigner functions, {\em Phys.~Rev.~A} {\bf 73} (2006) 012301.
\bibitem{Paz2} J.~P.~Paz, A.~J.~Roncaglia, and M.~Saraceno, Qubits in phase space: Wigner-function approach to quantum error correction and the mean king problem, {\em Phys.~Rev.~A} {\bf 72} (2005) 012309.
\bibitem{vanDam} W.~van Dam and M.~Howard, Noise Thresholds for Higher Dimensional Systems Using the Discrete Wigner Function, {\em Phys.~Rev.~A} {\bf 83} (2011) 032310.
\bibitem{Braasch1}  W.~F.~Braasch Jr.~and W.~K.~Wootters, A quantum prediction as a collection of epistemically restricted classical predictions, {\em Quantum} {\bf 6} (2022) 659.
\bibitem{Braasch2} W.~F.~Braasch Jr.~and W.~K.~Wootters, A classical interpretation of quantum theory? {\em Entropy} {\bf 24} (2022) 137.
\bibitem{Holevo1} A.~S.~Holevo, Bounds for the quantity of information transmitted by a quantum communication channel, {\em Problemy Peredachi Informatsii} {\bf 9} (1973) 3.
\bibitem{Holevo2} A.~S.~Holevo, Information-theoretical aspects of quantum measurement, {\em Problemy Peredachi Informatsii} {\bf 9} (1973) 31.
\bibitem{Holevo3} A.~S.~Holevo, Remarks on optimal quantum measurements, {\em Problemy Peredachi Informatsii} {\bf 10} (1974) 51.
\bibitem{Holevo4} A.~S.~Holevo, Problems in the mathematical theory of quantum communication channels, {\em Rep.~Math.~Phys.} {\bf 12} (1977) 273.
\bibitem{Holevo5} A.~S.~Holevo, Estimation of shift parameters of a quantum state, {\em Rep.~Math.~Phys.} {\bf 13} (1978) 379.  
\bibitem{Ivanovic} I.~D.~Ivanovic, Geometrical description of quantal state determination, {\em J.~Phys.~A} {\bf 14} (1981) 3241.
\bibitem{Fields} W.~K.~Wootters and B.~D.~Fields, Optimal state-determination by mutually unbiased measurements, {\em Ann.~Phys.~(N.Y.)} {\bf 191} (1989) 363.
\bibitem{Petz} D.~Petz, K.~M.~Hangos, and A.~Magnar, Point estimation of states of finite quantum systems, {\em J.~Phys.~A: Math.~Theor.} {\bf 40} (2007) 7955.
\bibitem{Bandyopadhyay} S.~Bandyopadhyay, P.~O.~Boykin, V.~Roychowdhury, and F.~Vatan, A New Proof of the Existence of Mutually Unbiased Bases, {\em Algorithmica} {\bf 34} (2002) 512.
\bibitem{Pittenger} A.~O.~Pittenger and M.~H.~Rubin, Mutually unbiased bases, generalized spin matrices and separability, {\em Linear Algebra and its Applications} {\bf 390} (2004) 255.
\bibitem{Chau} H.~F.~Chau, Unconditionally secure key distribution in higher dimensions by depolarization, {\em IEEE Trans.~Inf.~Theory} {\bf 51} (2005) 1451.
\bibitem{Sussman} W.~K.~Wootters and D.~M.~Sussman, Discrete phase space and minimum-uncertainty states, arxiv:0704.1277 (2007).
\bibitem{Calderbank} N.~Rengaswamy, R.~Calderbank, S.~Kadhe, and H.~D.~Pfister, Synthesis of Logical Clifford Operators via Symplectic Geometry, {\em Proc.~2018 IEEE International Symposium on Information Theory (ISIT)}, 791--795 (2018). 

%\bibitem{Leonhardt} U.~Leonhardt, Discrete Wigner functions and quantum-state tomography, {\em Phys.~Rev.~A} {\bf 53} (1996) 2998.  
\bibitem{ApplebyJMP} D.~M.~Appleby, Symmetric informationally complete--positive operator valued measures and the extended Clifford group, {\em J.~Math.~Phys.} {\bf 46} (2005) 052107.
\bibitem{Bertrand} J.~Bertrand and P.~Bertrand, A tomographic approach to Wigner's function, {\em Found.~Phys.} {\bf 17} (1987) 397.

\end{thebibliography}
\end{document}